\documentclass[showpacs,pra,aps,superscriptaddress,twocolumn,floatfix]{revtex4}
\usepackage[usenames,dvipsnames]{xcolor}
\usepackage{amsmath}
\usepackage{amssymb,mathrsfs}
\usepackage{graphicx}
\usepackage{mathtools}
\DeclareMathOperator{\e}{e}
\DeclareMathOperator{\eps}{\epsilon}
\DeclareMathOperator{\cc}{c.c.}

\usepackage{hyperref}
\allowdisplaybreaks

\begin{document}

\title{Nonlinear waves in coherently coupled Bose-Einstein
  condensates} 
\author{T. Congy}\affiliation{LPTMS, CNRS,
  Univ. Paris-Sud, Universit\'e Paris-Saclay, 91405 Orsay, France}
\author{A. M. Kamchatnov}\affiliation{Institute of Spectroscopy,
  Russian Academy of Sciences, Troitsk, Moscow, 108840, Russia}
\author{N. Pavloff}\affiliation{LPTMS, CNRS, Univ. Paris-Sud,
  Universit\'e Paris-Saclay, 91405 Orsay, France} \date{\today}

\begin{abstract}
  We consider a quasi-one-dimensional two-component Bose-Einstein
  condensate subject to a coherent coupling between its components,
  such as realized in spin-orbit coupled condensates. We study how
  nonlinearity modifies the dynamics of the elementary
  excitations. The spectrum has two branches which are affected in
  different ways. The upper branch experiences a modulational
  instability which is stabilized by a long wave-short wave resonance
  with the lower branch. The lower branch is stable.  In the limit of
  weak nonlinearity and small dispersion it is described by a
  Korteweg-de Vries equation or by the Gardner equation, depending on
  the value of the parameters of the system.
\end{abstract}


\maketitle

\section{Introduction}

The Bose-Einstein condensation of a mixture of different hyperfine
states of the same element (first realized by the JILA group
\cite{Wya97}) offers the possibility to transfer atoms from an
internal state to another one in a macroscopic matter wave. This
feature has driven a rich body of experimental studies of phenomena
such as the formation of spin domains, vortices and other nonlinear
structures \cite{domains}, internal Josephson effect \cite{Josephson},
formation of squeezed and entangled states \cite{intrication}, motion
of spin impurities \cite{impurity}, persistent currents \cite{Bea13},
effective gauge potentials \cite{gauge} and spin-orbit coupled systems
\cite{spin-orbit-exp}, which has itself opened an avenue of new
researches: observation of a superfluid Hall effect \cite{Leb12}; of
Zitterbewegung \cite{Qu13}; of spin Hall effect
\cite{Bee13}; of tunable Landau-Zener transitions \cite{Ols14}; of a
Dicke-type phase transition \cite{Ham14}; of the softening of a
roton-like dispersion relation \cite{roton}...

In some of the above cited works, the change of internal state is
only due to spin-dependent collisions, but in others the coupling is
externally driven by a combination of radiofrequency and microwave
fields \cite{two1} or by using Raman coupling lasers \cite{two2}. In
the present study we concentrate on an effective spin $1/2$ system in
which two internal states are coherently coupled by an external
potential. In such a system, the coupling explicitly breaks the
$U(1)\times U(1)$ symmetry originating from the irrelevance of global
phase factors of each the two components: the relative phase is no
longer free and only remains a $U(1)$ symmetry for the global phase of
the spinor. As a consequence, the two-branched spectrum of the system
has a single Goldstone mode and the other branch is gaped. The mean
field dynamics of the system is described by two coupled
Gross-Pitaevskii equations accounting for intra and inter-species
collisions, for the external coupling field and also possibly for a
spin-orbit term. The ground state of the system and the associated
possible phase transitions and the elementary excitations have been
theoretically studied in Refs.~\cite{GS1} and \cite{GS2}, as long as a
rich variety of nonlinear structures (Refs. \cite{NL1} and
\cite{NL2}).

The reason for the protean aspect of the theoretical approaches of the
system lies in the fact that its dynamics is described by a
nonintegrable set of coupled Gross-Pitaevskii equations which do not
admit simple integrable equations as limiting cases. Even in the
simpler case of a spinor condensate in the absence of spin-orbit and
Raman coupling, the integrable limit is the so-called Manakov system
(obtained when all the nonlinear interaction constants are equal)
which does not pertain to the well-studied Ablowitz-Kaup-Newell-Segur
hierarchy and for which all the types of solutions are not yet fully
classified (see e.g., Refs.~\cite{Kam15}). The aim of the present work
is to partially clarify the rich nonlinear behavior of the system by
presenting a systematic study revealing how nonlinear effects modify
the elementary excitations of the system.

The paper is organized as follows: the model, its ground state and
linear excitations are described in Sec.~\ref{model}. We then use a
singular perturbation theory to describe in Section
\ref{sec:upperbranch} how excitations in the upper branch of the
dispersion relation are affected by nonlinear effects. The method is
exposed in subsections \ref{subeps1}, \ref{sec:order2} and
\ref{order3} and the results are summarized and discussed in
subsection \ref{final-upper}. The technique used in
Sec.~\ref{sec:upperbranch} can also be employed for describing the
effects of nonlinearity on the lower branch of the spectrum.  However,
for this branch another approach can be used which is more appropriate
in the long wave length limit.  This is explained in
Sec.~\ref{sec:lowerbranch} and we show in subsections \ref{KdVregime}
and \ref{gardner_regime} how to deal with this issue. The general
doctrine is presented in subsection \ref{quartic} where we also
discuss the different regimes accessible in present days experiments.
Our conclusions are summarized in Sec.~\ref{conclusion} and some
technical aspects are detailed in Appendices \ref{app1} and
\ref{app2}.

\section{The model and elementary excitations}\label{model}

We consider a one-dimensional system described by a two-component
spinor order parameter $\Psi(x,t)=(\psi_\uparrow,\psi_\downarrow)^t$
(where the superscript $^t$ denotes the transposition) obeying the
following coupled Gross-Pitaevskii equations
\begin{equation}\label{mod1a}
i\,\hbar\, \partial_t \Psi
=H_0 \, \Psi + \begin{pmatrix}
\alpha_1 |\psi_\uparrow|^2    & &\alpha_2 \psi_\downarrow^*\psi_\uparrow \\[4 mm]
\alpha_2 \psi_\uparrow^*\psi_\downarrow & &\alpha_1 |\psi_\downarrow|^2 \\
\end{pmatrix}
\Psi\; ,
\end{equation}
where $H_0$ is the single particle Hamiltonian:
\begin{equation}\label{mod1b}
H_0=
\frac{1}{2 m}\left(\frac{\hbar}{i}\,\partial_x-\hbar\, k_0\, \sigma_z\right)^2
+\frac{\hbar\Omega}{2}\,\sigma_x  \; ,
\end{equation}
$\sigma_x$ and $\sigma_z$ being Pauli matrices. This corresponds to a
system with equal contribution of Rashba and Dresselhaus coupling, as
realized in spin-orbit coupled condensates (see, e.g.,
Ref.~\cite{Li14}). In Eq.~\eqref{mod1a}, $\alpha_2 =
\alpha_{\uparrow\downarrow}$ is the interspecies interaction
coefficient, and for simplicity we have assumed equal intraspecies
interaction: $\alpha_{\uparrow\uparrow}=
\alpha_{\downarrow\downarrow}\equiv\alpha_1$. In the following we will
consider the case of repulsive intraspecies interaction: $\alpha_1>0$.

It is convenient to
re-parametrize the spinor wave-function \cite{ktu-2005} :
\begin{equation}\label{mod2}
\Psi(x,t)=\begin{pmatrix}
\psi_\uparrow \\
\psi_\downarrow \\
\end{pmatrix}
=
\sqrt{\rho}\, \e^{i\Phi/2}
\begin{pmatrix}
\cos\frac{\theta}2\,\e^{-i\varphi/2} \\[1mm]
\sin\frac{\theta}2\,\e^{i\varphi/2}  \\
\end{pmatrix}\; .
\end{equation}
Here $\rho(x,t)=|\psi_\uparrow|^2+|\psi_\downarrow|^2$ denotes the total
density of the condensate and $\Phi(x,t)$ has the meaning of the
velocity potential of its in-phase motion; the angle $\theta(x,t)$ is
the variable describing the relative density of the two components
($\cos\theta=(|\psi_\uparrow|^2-|\psi_\downarrow|^2)/\rho$) and the phase
$\varphi(x,t)$ is the potential of their relative (counter-phase)
motion. Accordingly, the densities of the components of the condensate
are given by
\begin{equation}\label{mod3}
\begin{split}
    \rho_\uparrow(x,t)=& |\psi_\uparrow|^2=\rho\cos^2(\theta/2)\; ,\\
\rho_\downarrow(x,t)=& |\psi_\downarrow|^2=\rho\sin^2(\theta/2)\; .
\end{split}
\end{equation}
Their velocities are defined as
\begin{equation}\label{mod4}
\begin{split}
v_{\uparrow}(x,t)=& \tfrac12(\Phi_x - \varphi_x)-k_0\; ,\\
v_{\downarrow}(x,t)=& \tfrac12(\Phi_x+ \varphi_x)+k_0\; .
\end{split}
\end{equation}
It will also appear convenient to define the following
velocity fields
\begin{equation}\label{kdv0}
U(x,t)=\Phi_x\; , \quad\mbox{and}\quad v(x,t)=\varphi_x\, .
\end{equation}

Equation \eqref{mod1a}, expressed in terms of the real fields $\Phi$, $\rho$,
$\theta$ and $\varphi$, reads
\begin{subequations}\label{mod5}
    \begin{align}
    \rho_t=&\tfrac12[\rho\,(\varphi_x+2k_0)\cos\theta]_x-
\tfrac12(\rho\,\Phi_x)_x \; ,\\
    - \Phi_t=&- \tfrac12 \cot\theta\frac{(\rho\,\theta_x)_x}{\rho}
+ \tfrac12 \left(\frac{\rho_x^2}{2\rho^2}-\frac{\rho_{xx}}{\rho}\right)
+\nonumber \\
&    \tfrac14[\Phi_x^2+\theta_x^2+(\varphi_x+2k_0)^2]+\nonumber \\
&(\alpha_1+\alpha_2)\rho+\Omega\, \frac{\cos\varphi}{\sin\theta} \; ,
\label{mod5b} \\
-\theta_t=&\tfrac12 \Phi_x\theta_x +
\frac{1}{2\rho} [\rho(\varphi_x+2 k_0)\sin\theta]_x
+\nonumber \\
 & \Omega\,\sin\varphi \; ,\\
    \varphi_t=&\frac{1}{2\sin\theta}\frac{(\rho\,\theta_x)_x}{\rho}
-\tfrac12 \Phi_x(\varphi_x+2k_0)+\nonumber \\
& (\alpha_1-\alpha_2)\rho \cos\theta-\Omega\,\cos\varphi\cot\theta \; ,
\label{mod5d}    \end{align}
\end{subequations}
where we have used units such that $\hbar=1=m$.

In all the following we will assume that the different parameters of
the Hamiltonian are fixed in such a way that the ground state of the
system corresponds to a configuration in which both components are
homogeneous ($\rho=\rho_0$ and $\theta=\theta_0$), in phase
($\varphi=0$), stationary ($\rho$, $\Phi_x$, $\theta$ and $\varphi_x$ are
time independent) with equal densities ($\theta_0=-\pi/2$
\cite{rem_theta}). In this case, one obtains $\Phi=-2\,\mu \,t$, where
\begin{equation}\label{mod6}
\mu=\frac{k_0^2}{2}+\frac{g_1}{2} -\frac{\Omega}{2}
\end{equation}
is the chemical potential. In this expression we have used the notation
$g_1=(\alpha_1+\alpha_2)\rho_0$. It will also appear convenient to
define $g_2=(\alpha_1-\alpha_2)\rho_0$ and to introduce a rescaled
density $n(x,t)=\rho(x,t)/\rho_0$.

\begin{figure}
\includegraphics[width=0.99\linewidth]{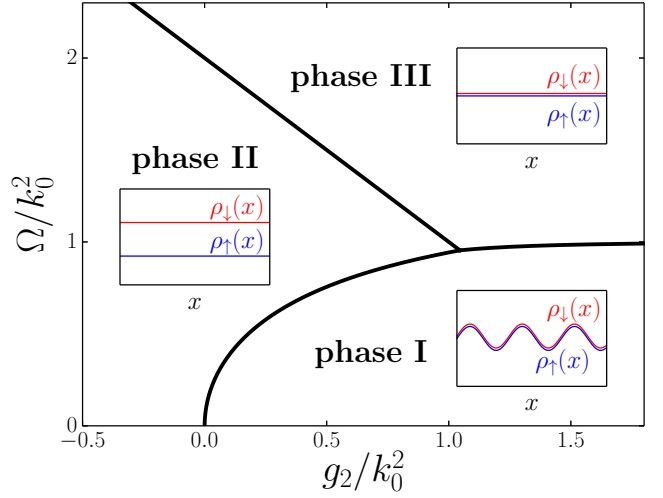}
\caption{(Color online) Schematic phase space of the spin orbit
  coupled system as a function of the parameters $\Omega/k_0^2$ and
  $g_2/k_0^2$. For each phase the inset represent a typical density
  pattern. The boundary between phases III and II corresponds to
  $\Omega+g_2=2 \, k_0^2$.}\label{fig.phases}
\end{figure}

In the absence of spin-orbit coupling ($k_0=0$) this ground state is
stable provided $\Omega+g_2>0$ \cite{GS1}. For a spin-orbit coupled
system, this ground state is denoted as the ``single minimum'' or
``zero momentum'' or ``phase III'' ground state. It is the true ground
state of the system in a region of parameters which is schematically
depicted in Fig.~\ref{fig.phases} (adapted from Ref.~\cite{Li14}).

Although the present work is devoted to the study of nonlinear effects
in phase III, we note that the methods we use also apply---with
unessential modifications---in phase II which is a spin polarized
phase where the system condensates in a single plane wave state with
non-zero momentum. Phase I (the so-called
striped phase) which has a modulated ground state density deserves a
special treatment.

\begin{figure}
\includegraphics[width=0.99\linewidth]{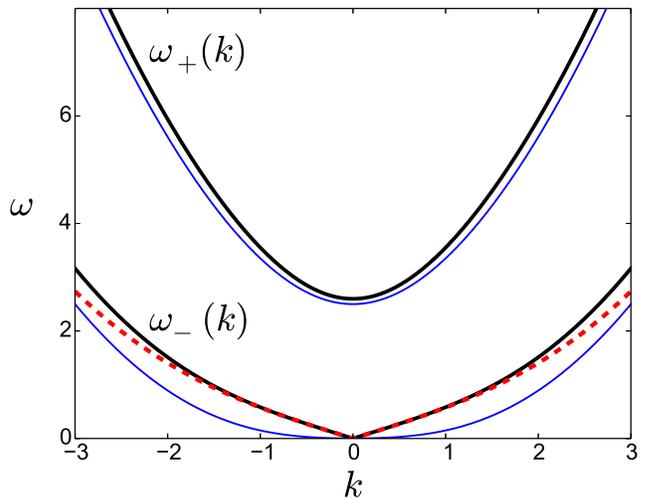}
\caption{(Color online) The black solid lines represent the exact
  dispersion relations $\omega_+(k)$ (upper branch) and $\omega_-(k)$
  (lower branch), solutions of Eq. \eqref{linear3}. The Figure is
  drawn in the case $\alpha_1\rho_0=1.2$, $\alpha_2\rho_0=1.0$,
  $k_0=1.0$ and $\Omega=2.5$. The (red) dashed line represent the long
  wavelength expansion \eqref{sol:disp}. The thin (blue) lines
  represent the spectrum of the single particle Hamiltonian $H_0$
  [cf. Eq.~\eqref{mod1b}].  They are obtained by taking $g_1=g_2=0$ in
  Eq. \eqref{linear3}: in this case one obtains
  $\omega_{\pm}(k)=k^2/2+\Omega/2\pm [k_0^2 k^2 + \Omega^2/4]^{1/2}$. }
\label{fig.dispers}
\end{figure}

A first insight in the dynamics of the system can be obtained by
linearizing Eqs.~\eqref{mod5}. For simplifying the notations we
introduce the column vector
\begin{equation}\label{mod7}
\Xi(x,t)=\begin{pmatrix} n \\ \Phi \\ \theta\\ \varphi\end{pmatrix} \; ,
\quad\mbox{with}\quad
\Xi^{(0)}(t)=\begin{pmatrix} 1 \\ -2\mu t \\ -\pi/2 \\ 0\end{pmatrix}
\end{equation}
being the ground state value of $\Xi(x,t)$.
We write
\begin{equation}\label{mod7bis}
\Xi(x,t)=\Xi^{(0)}(x,t)+\Xi'(x,t)\; ,
\end{equation}
where $\Xi'(x,t)$ describes a small departure of the fields $n$,
$\Phi$, $\theta$ and $\varphi$ from their ground state
values. Inserting this ansatz into \eqref{mod5} one obtains at first
order in $\Xi'$ a system of the form
\begin{equation}\label{mod8}
\mathbb{M}(\partial_x,\partial_t) \, {\Xi}' =0\; ,
\end{equation}
where
\begin{equation}\label{mod9}
\mathbb{M}= \begin{pmatrix}
\partial_t & \frac{\partial^2_x}{2} & -k_0\partial_x & 0 \\
-\frac{\partial_x^2}{2}+g_1 & \partial_t & 0 & k_0\partial_x \\
-k_0\partial_x & 0  & \partial_t & -\frac{\partial^2_x}{2}+\Omega \\
0 & k_0\partial_x & \frac{\partial^2_x}{2}-\Omega-g_2 & \partial_t \end{pmatrix}
\; .
\end{equation}
This equation being linear one can expand $\Xi'(x,t)$ on a basis of
plane waves of wave-vector $k$ and angular frequency $\omega$.  This
amounts to look for solutions of \eqref{mod8} of the form
$\Xi'(x,t)=\hat{\Xi}'\exp[i(k x -\omega t)] + \cc$, where ``$\cc$''
stands for ``complex conjugate'' and $\hat{\Xi}'$ is a constant co\-lumn
vector whose entries are possibly complex. One then obtains a system of
linear equations which reads
\begin{equation}\label{linear2a}
\mathbb{M}_1\, \hat{\Xi}'=0\; ,
\end{equation}
where
\begin{equation}\label{linear2}
\begin{split}
\mathbb{M}_1=& \mathbb{M}(i k,-i\omega) \\
=&
\begin{pmatrix}
-i\omega & -\frac{k^2}{2} & -i k_0 k & 0 \\
\frac{k^2}{2}+g_1 & -i\omega & 0 & i k_0 k \\
-i k_0 k &  0 & -i\omega & \frac{k^2}{2}+\Omega \\
0 & i k_0 k & -\frac{k^2}{2}-\Omega-g_2 & -i\omega \end{pmatrix}
\; .
\end{split}
\end{equation}
The system \eqref{linear2a} has non-trivial solutions only if the
determinant of $\mathbb{M}_1$ vanishes. This fixes the dispersion
relation of the elementary excitations, with two branches
$\omega=\omega_\pm(k)$ which are represented in
Fig.~\ref{fig.dispers}. They are solutions of
\begin{equation}\label{linear3}
\begin{split}
0& = \omega^4
 -\omega^2\bigg[\frac{k^4}{2}+2 k_0^2 k^2 +\Omega k^2\\
& +\Omega^2
+(g_1+g_2)\frac{k^2}{2}+\Omega g_2\bigg]\\
&+\frac{k^2}{2}
\left[\frac{k^2}{2}+\Omega+g_2-2 k_0^2\right]\times
\\
&
\left[\left(\frac{k^2}{2}+\Omega\right)
\left(\frac{k^2}{2}+g_1\right)-k^2 k_0^2\right]\; .
\end{split}
\end{equation}

The upper branch $\omega=\omega_+(k)$ is gaped, with a dispersion
relation of the form
\begin{equation}\label{upper-disp}
\omega_+(k)=\sqrt{\Omega(\Omega+g_2)} + {\cal O}(k^2) \; .
\end{equation}
The lower branch $\omega=\omega_-(k)$ is not gaped: it accounts for
the Goldstone mode corresponding to the spontaneous breaking of the
global $U(1)$ symmetry of the system. One sees in
Fig.~\ref{fig.dispers} that the upper branch is not qualitatively
affected by interaction effects, contrarily to the lower branch whose
long wavelength dispersion relation would be quadratic in the absence
of interaction and becomes linear in its presence. The lower branch
admits, for the positive $k$ portion of the spectrum, the following
expansion (corresponding to linear waves propagating in the
positive-$x$ direction):
\begin{equation}\label{sol:disp}
\omega_-(k) = c\,k+c_3\,k^3+{\cal O}(k^5)\; ,
\end{equation}
where
\begin{equation}\label{eq:sound_v}
c =\sqrt{\frac{g_1}{2}
\left(1-\frac{2k_0^2}{\Omega+g_2}\right)}\;,
\end{equation}
is the sound velocity, and the parameter $c_3$ verifies
\begin{equation}\label{eq:c3bis}
\begin{split}
	4 \,c \,c_3 =\;& \frac{1}{2} -
\frac{2 k_0^2}{\Omega(\Omega+g_2)}
\bigg[2\Omega+g_1+g_2 \\
	& \quad-
\frac{(\Omega+g_1+g_2)(2 \Omega-g_1+g_2)}{2(\Omega+g_2)}
\\
	& \quad-k_0^2
\frac{(\Omega+g_1+g_2)^2}{(\Omega+g_2)^2}
\bigg]\;.
\end{split}
\end{equation}

\section{Nonlinear perturbation theory for excitations
propagating in the upper branch}\label{sec:upperbranch}

We study in the present section how nonlinear effects modify the
structure of an excitation propagating in the upper branch of the
spectrum. For instance, one can anticipate that nonlinear terms cause
some modulations or anharmonicities of this wave, and make it interact
with the other branch of the spectrum. Instead of the simple linear
analysis of section \ref{model} (Eqs.~\eqref{mod7}, \eqref{mod7bis}
and following), we perform here a singular perturbative expansion by
writing the term $\Xi'(x,t)$ in Eq.~\eqref{mod7bis} under the form
(see, e.g., \cite{Jef82,Tan83,New85,Kam00,Abl11}) :
\begin{equation}\label{ansatz_upper}
\Xi'(x,t)= \sum_{n\geq1} \epsilon^n \, \Xi^{(n)}(x,t,X,T_1,T_2)\; .
\end{equation}
In this expansion $\epsilon$ is a small parameter.
\begin{equation}\label{multiscale0}
X=\epsilon x \quad\mbox{and}\quad
T_2=\epsilon T_1=\epsilon^2 t \; ,
\end{equation}
are multiscale coordinates aiming at
describing the slow spatial and temporal modulations of a wave packet
of finite amplitude.

$\Xi^{(0)}$ in \eqref{ansatz_upper} is the same as in \eqref{mod7} and
we make the following ansatz for the form of the ${\cal O}(\eps)$
term:
\begin{equation}\label{init-ansatz}
\begin{split}
\Xi^{(1)}(x,t,X,T_1,T_2)=\;& \overline{\Xi}_0^{(1)}(X,T_1,T_2) \\
+&
\left(\widetilde\Xi_1^{(1)}(X,T_1,T_2) \e^{i \beta(x,t)} + \cc\right)
\; ,
\end{split}
\end{equation}
where
\begin{equation}\label{beta}
\beta(x,t)=k x -\omega t \; .
\end{equation}
This means that we assume that the ${\cal O}(\epsilon)$ solution of
\eqref{mod5} consists in a slowly varying contribution
($\overline{\Xi}_0^{(1)}$, \cite{Rembar}) plus an oscillating term
with a smoothly varying amplitude $\widetilde\Xi_1^{(1)}$. We will see
below that the non-oscillating contribution $\overline{\Xi}_0^{(1)}$
is necessary for the consistency of the approach, meaning that
nonlinearity not only modifies the shape of a finite amplitude wave
but also affects the background on top of which the wave propagates.

We enforce a behavior of type \eqref{init-ansatz} only at order
$\eps$; it then will be automatically verified at higher orders, with
also possible contributions from higher harmonics: see \eqref{eq:order2_ansatz}
for the form of the ${\cal O}(\eps^2)$ solution.

The multiscale analysis consists in considering that the time
variables $t$, $T_1$ and $T_2$ (and also the spatial coordinates $x$
and $X$) are independent. One thus writes
\begin{equation}\label{derivatives}
\begin{split}
& \partial_x= k\, \partial_\beta + \eps \partial_X \; ,\\
\mbox{and}\qquad & \partial_t= - \omega\, \partial_\beta
+ \eps \partial_{T_1}+\eps^2\partial_{T_2} \;.
\end{split}
\end{equation}
The method applies for any value of $k_0$, provided one remains in
phase III, but the general expressions are quite cumbersome: For
legibility we present the computation in the simpler case
$k_0=0$.

\subsection{Order $\eps$}\label{subeps1}

At this order, Eq.~\eqref{mod5} reads [as already obtained in
Eq.~\eqref{mod8}]
\begin{equation}\label{order1}
\mathbb{M}{(k\partial_\beta,-\omega\partial_\beta)} \, \Xi^{(1)} = 0
\; ,
\end{equation}
where $\mathbb{M}$ is defined in \eqref{mod9}.  Using the matrix
$\mathbb{M}_1$ of Eq.~\eqref{linear2} and defining $\mathbb{M}_0$ by
\begin{equation}
\label{eq:M0}
\mathbb{M}_0 = \mathbb{M}\Big(0,0\Big)  =
\begin{pmatrix}
0 & 0 & 0 & 0\\
g_1 & 0 & 0 & 0 \\
0 & 0 & 0 & \Omega \\
0 & 0 & -\Omega-g_2 & 0
\end{pmatrix} \;,
\end{equation}
one can re-write Eq.~\eqref{order1} as
\begin{equation}\label{order1b}
\begin{split}
\mathbb{M}_0 \; \overline\Xi_0^{(1)}(X,T_1,T_2)= 0 & \; ,\\
\mbox{and}\quad \mathbb{M}_1\; \widetilde\Xi_1^{(1)}(X,T_1,T_2)=
0 &\; .
\end{split}
\end{equation}
For \eqref{order1b} to have nontrivial solutions we need to impose
$\det \mathbb{M}_1 = 0$ (we already have $\det
\mathbb{M}_0 = 0$). As was seen in section \ref{model}, this determines
the dispersion relation. We study here a wave propagating in the upper
branch of the spectrum, that is, in the expression \eqref{beta} for
$\beta(x,t)$, one has $\omega=\omega_+(k)$. We then obtain for the
solutions $\overline\Xi^{(1)}_0$ and $\widetilde\Xi^{(1)}_1$ of
Eqs.~\eqref{order1b} expressions of the form:
\begin{equation}\label{order1c}
\begin{split}
\overline\Xi^{(1)}_0(X,T_1,T_2) = &
\begin{pmatrix}
\overline{n}^{(1)}\\ \overline\Phi^{(1)} \\
\overline\theta^{(1)}\\ \overline\varphi^{(1)}
\end{pmatrix} =
\begin{pmatrix}
0 \\ 1 \\ 0 \\ 0
\end{pmatrix}
\overline\Phi^{(1)} \\
\equiv & \overline{R}_0\, \overline\Phi^{(1)}(X,T_1,T_2)
\; ,
\end{split}
\end{equation}
and
\begin{equation}\label{order1d}
\begin{split}
\widetilde\Xi_1^{(1)}(X,T_1,T_2)= &
\begin{pmatrix}
\widetilde{n}^{(1)}\\ \widetilde\Phi^{(1)}\\
\widetilde\theta^{(1)}\\ \widetilde\varphi^{(1)}
\end{pmatrix}
=
\begin{pmatrix}
0 \\ 0 \\ 1 \\ i\, \Delta
\end{pmatrix}
\widetilde\theta^{(1)}\\
\equiv & \widetilde{R}_1 \, \widetilde\theta^{(1)}(X,T_1,T_2) \; ,
\end{split}
\end{equation}
where $\Delta=(1+\frac{2\,g_2}{k^2+2\,\Omega})^{1/2}$.  At this point
$\overline\Phi^{(1)}(X,T_1,T_2)$ and
$\widetilde\theta^{(1)}(X,T_1,T_2)$ in expressions \eqref{order1c} and
\eqref{order1d} are still unknown, but we already collected some
useful pieces of information on the form of the wave: we see that
$\overline{n}^{(1)} =\overline\theta^{(1)} =\overline\varphi^{(1)}
=\widetilde{n}^{(1)} =\widetilde\Phi^{(1)} =0$ and that
$\widetilde\varphi^{(1)}$ is proportional to $\widetilde\theta^{(1)}$.
In the case $k_0\neq 0$, $\overline{n}^{(1)}$ and
$\widetilde\Phi^{(1)}$ are non zero, but both are proportional to
$\widetilde\theta^{(1)}$, as well as $\widetilde\varphi^{(1)}$.

\subsection{Order $\eps^2$}\label{sec:order2}

At this order Eq.~\eqref{mod5} reads
\begin{equation}
\label{eq:order2}
\begin{split}
& \mathbb{M}(k \partial_\beta,-\omega \partial_\beta)
\left[\Xi^{(2)}(x,t,X,T_1,T_2)\right]=\\
& \overline{C}_0(X,T_1,T_2) + \\
& \left[\widetilde{C}_1(X,T_1,T_2) \e^{i \beta(x,t)}+\cc
\right] + \\
& \left[\widetilde{C}_2(X,T_1,T_2) \e^{2 i \beta(x,t)}+ \cc \right]
\; .
\end{split}
\end{equation}
with
\begin{equation}\label{lesc0}
\overline{C}_0(X,T_1,T_2) =
\begin{pmatrix}
0 \\ 1 \\ 0 \\ 0
\end{pmatrix}\partial_{T_1} \overline{\Phi}^{(1)} +
\begin{pmatrix}
0 \\ g_2 \\ 0 \\ 0
\end{pmatrix} \left|\widetilde{\theta}^{(1)}\right|^2 \;,
\end{equation}
\begin{equation}\label{lesc1}
\widetilde{C}_1(X,T_1,T_2) =
\begin{pmatrix}
0 \\ 0 \\ 1 \\ i\,\Delta
\end{pmatrix} \partial_{T_1}\widetilde\theta^{(1)} \\
+
\begin{pmatrix}
0 \\ 0 \\ \Delta \, k \\ i\,k
\end{pmatrix} \partial_{X}\widetilde\theta^{(1)} \;,
\end{equation}
and
\begin{equation}\label{lesc2}
\widetilde{C}_2(X,T_1,T_2) =
\begin{pmatrix}
i\,\Delta\, k^2  \\ \frac{1}{2} \left(
\frac{k^2-2\,\Omega}{k^2+2\,\Omega}\,g_2 - k^2 - 2\,\Omega \right) \\ 0 \\ 0
\end{pmatrix}(\widetilde\theta^{(1)})^2\; .
\end{equation}
In expressions \eqref{lesc1} and \eqref{lesc2} we have used the
same notation $\Delta$ as in \eqref{order1d}.
The precise expressions \eqref{lesc0}, \eqref{lesc1} and \eqref{lesc2}
for $\overline{C}_0$, $\widetilde{C}_1$ and $\widetilde{C}_2$ result
from the formulas \eqref{order1c} and \eqref{order1d} for
$\overline\Xi^{(1)}_0$ and $\widetilde\Xi_1^{(1)}$.

Since the operator $\mathbb{M}(\partial_x,\partial_t)$
is linear, the solution of equation \eqref{eq:order2}
consists of three contributions, one for each of the source terms. Hence
$\Xi^{(2)}$ is of the form
\begin{equation}\label{eq:order2_ansatz}
\begin{split}
\Xi^{(2)}& = \overline\Xi^{(2)}_0(X,T_1,T_2)\\
&+ \left[\widetilde\Xi^{(2)}_1(X,T_1,T_2)\e^{i\beta} + \cc\right]\\
&+ \left[\widetilde\Xi^{(2)}_2(X,T_1,T_2)\e^{2i\beta} + \cc\right]\;,
\end{split}
\end{equation}
the different components being solutions of
\begin{equation}\label{eq:order2_0}
\mathbb{M}_0 \; \overline\Xi^{(2)}_0(X,T_1,T_2) =
\overline{C}_0(X,T_1,T_2)\; ,
\end{equation}
\begin{equation}\label{eq:order2_1}
\mathbb{M}_1\; \widetilde\Xi^{(2)}_1(X,T_1,T_2) =
\widetilde{C}_1(X,T_1,T_2)\; ,
\end{equation}
and
\begin{equation}\label{eq:order2_2}
\mathbb{M}_2 \; \widetilde\Xi^{(2)}_2(X,T_1,T_2) =
\widetilde{C}_2(X,T_1,T_2)\; ;
\end{equation}
where $\mathbb{M}_2$ is defined similarly to $\mathbb{M}_1$ in
Eq.~\eqref{linear2} and $\mathbb{M}_0$ in
Eq.~\eqref{eq:M0}:
\begin{equation}\label{def:M2}
\mathbb{M}_2 =
\mathbb{M}\Big( 2i k, -2i \omega_+(k) \Big)\; .
\end{equation}
Equation \eqref{eq:order2_2} is easily solved because $\det
\mathbb{M}_2 \neq 0$ \cite{remshg}. We do not write its solution
explicitly, but it is necessary for next order in $\eps$: it
contributes to the r.h.s. of Eq. \eqref{eq:order3}, in particularly to
the expression \eqref{lesd1} for the coefficient $\widetilde{D}_1$.

Solving Eqs.~\eqref{eq:order2_0} and \eqref{eq:order2_1} is more
complicated than solving Eq.~\eqref{eq:order2_2} because
$\det \mathbb{M}_0=0$ and $\det \mathbb{M}_1=0$.
Hence, if $\overline{C}_0$ is not in the image space of $\mathbb{M}_0$
(or if $\widetilde{C}_1$ is not in the image space of $\mathbb{M}_1$)
one cannot find a solution. One must thus impose that $\overline{C}_0$
is in the image space of $\mathbb{M}_0$ and that $\widetilde{C}_1$ is
in the image space of $\mathbb{M}_1$. This can be done conveniently through the
following technique. Let us define
$\overline{L}_0$ and $\widetilde{L}_1$ such that
\begin{equation}\label{consistant}
\begin{split}
& \mathbb{M}_0^t \, \overline{L}_0=0 \quad \Rightarrow \quad
\overline{L}_0 =
\begin{pmatrix} 1\\0\\0\\0\end{pmatrix}\; ,
\\
& \mathbb{M}_1^t\, \widetilde{L}_1=0 \quad \Rightarrow \quad
\widetilde{L}_1 =
\begin{pmatrix}0\\0\\1\\-i/\Delta \end{pmatrix}
\; .
\end{split}
\end{equation}
Multiplying \eqref{eq:order2_0} by the transposed row vector
$\overline{L}_0^{\, t}$
and \eqref{eq:order2_1} by $\widetilde{L}_1^{\, t}$ one obtains \cite{rem1}
\begin{equation}\label{Lorder2}
\begin{split}
\overline{L}_0^{\, t} \cdot \overline{C}_0 (X,T_1,T_2) =0 & \; , \\
\mbox{and}\quad
\widetilde{L}_1^{\, t} \cdot \widetilde{C}_1 (X,T_1,T_2) = 0 & \; .
\end{split}
\end{equation}
The first of these equations is trivially satisfied. The second
imposes that
\begin{equation}\label{eq-importante}
\partial_{T_1} \widetilde\theta^{(1)} + \omega_+^\prime(k)
\partial_X \widetilde\theta^{(1)} = 0
\; ,
\end{equation}
which implies the important physical result that the envelope of the
wave packet propagates with the group velocity
$\omega_+^\prime(k)={\rm d}\omega_+/{\rm d}k$.

Once Eq.~\eqref{eq-importante} is satisfied, the compatibility condition
\eqref{Lorder2} is fulfilled and
one can solve
Eqs.~\eqref{eq:order2_0} and \eqref{eq:order2_1}. One obtains
\begin{equation}\label{eq:solve2_0}
\overline\Xi^{(2)}_0(X,T_1,T_2) =
\begin{pmatrix}
\frac{1}{g_1}\partial_{T_1} \overline{\Phi}^{(1)}+\frac{g_2}{g_1}
\left|\widetilde{\theta}^{(1)}\right|^2  \\ 0 \\ 0 \\ 0
\end{pmatrix}
 \; ,
\end{equation}
and
\begin{equation}\label{eq:solve2_1}
\widetilde\Xi^{(2)}_1(X,T_1,T_2) =
\begin{pmatrix}
0 \\ 0 \\
\frac{i\,k\,g_2}{\omega_+^2(k)} \partial_X
\widetilde\theta^{(1)} \\ 0
\end{pmatrix}\\
\; .
\end{equation}

\subsection{Order $\eps^3$}\label{order3}

At this order one obtains an equation whose form is quite
similar to that of Eq.~\eqref{eq:order2} with additional harmo\-nics:
\begin{equation}
\label{eq:order3}
\begin{split}
& \mathbb{M}(k \partial_\beta,-\omega \partial_\beta)
\left[\Xi^{(3)}(x,t,X,T_1,T_2)\right]= \\
& \overline{D}_0(X,T_1,T_2) +\\
& \left[\widetilde{D}_1(X,T_1,T_2)  \e^{i \beta(x,t)}+\cc \right] +\\
& \left[\widetilde{D}_2(X,T_1,T_2)  \e^{2 i \beta(x,t)}+ \cc \right] +\\
& \left[\widetilde{D}_3(X,T_1,T_2)  \e^{3 i \beta(x,t)}+ \cc \right]\; .
\end{split}
\end{equation}
We need not write the expressions for $\widetilde{D}_2$ and
$\widetilde{D}_3$ because they are not necessary to determine the
dynamic of $\widetilde{\theta}^{(1)}$. The terms $\overline{D}_0$ and
$\widetilde{D}_1$ read (remember that for legibility we give the explicit
expressions only in the case $k_0=0$)
\begin{equation}\label{lesd0}
\begin{split}
\overline{D}_0 & =
\begin{pmatrix}
\frac{k \left( g_1(k^2+2\,\Omega+2\,g_2) + g_2(k^2+2\,\Omega+g_2)
\right)}{2\,g_1\,\omega_+(k)}
\\
\frac{i\,g_2^2\,k}{2\,\omega_+^2(k)}
\\ 0 \\ 0
\end{pmatrix}  \partial_X \left| \widetilde\theta^{(1)} \right|^2
\\
& +
\begin{pmatrix}
\frac{1}{2}\, \partial_X^2 \overline{\Phi}^{(1)} -
\frac{1}{g_1} \partial_{T_1}^2 \overline{\Phi}^{(1)}\\
\partial_{T_2} \overline{\Phi}^{(1)}\\ 0 \\ 0
\end{pmatrix}
\;,
\end{split}
\end{equation}
and
\begin{equation}\label{lesd1}
\begin{split}
\widetilde{D}_1&=
\begin{pmatrix}
0\\0 \\1 \\ i \Delta
\end{pmatrix}\partial_{T_2} \widetilde{\theta}^{(1)}
 +
\begin{pmatrix}
0\\0\\i\,P(k) \\ Q(k)
\end{pmatrix}\left|\widetilde\theta^{(1)}\right|^2\widetilde\theta^{(1)}
\\
& +
\begin{pmatrix}
0\\0\\
 - i \,\frac{8\, g_2\, \Omega \left(g_2+k^2+2\, \Omega \right)+\left(k^2+2\,
\Omega \right)^3}{16\, \omega_+^3(k)} \\
  \frac{1}{2} + \frac{k^2 g_2}{2\,
\omega_+^2(k)}
\end{pmatrix}\partial_X^2 \widetilde\theta^{(1)} \\
&+
\begin{pmatrix}
0\\0\\\frac{i\,k}{2}\\
-\frac{\Delta\,  k}{2}
\end{pmatrix} \widetilde\theta^{(1)}\partial_X \overline{\Phi}^{(1)}+
\begin{pmatrix}
0\\0\\0 \\\frac{g_2}{g_1}
\end{pmatrix}\widetilde\theta^{(1)}\partial_{T_1} \overline{\Phi}^{(1)}
 \; .
\end{split}
\end{equation}
In the above expression for $\widetilde{D}_1$ the quantities $P(k)$ and
$Q(k)$ are defined as
\begin{equation}
\begin{split}
P(k)=-&\frac{\sqrt{2 g_2+k^2+2 \Omega }}{4 \left(k^2+2 \Omega
\right)^{3/2} F(k)} \, \times\\
&\Big[\left(k^2+2 \Omega \right)
4 g_2 \left(-2 g_2 \Omega +k^4-4 k^2 \Omega -4 \Omega
^2\right)\\
& +\left(k^2+2 \Omega \right)^2 \left(3 k^2+2 \Omega \right)
\left(k^2-2 \Omega \right) \\
 &+ 2 g_1 k^2 \left(4 g_2
\Omega +\left(k^2+2 \Omega \right)^2\right) \Big]\; ,
\end{split}
\end{equation}
and
\begin{equation}
\begin{split}
Q(k)=&\frac{1}{4 g_1 \left(k^2+2 \Omega
\right) F(k)} \, \times \\
&\Big[-8 g_2^3 \left(k^2+2 \Omega \right)^2\\
& -g_1 \left(k^2+2
\Omega \right)^2 \left(2 g_1 k^2+\left(3 k^2+2 \Omega \right)
\left(k^2-2 \Omega \right)\right)\\
&+4 g_2^2 g_1 \left(5 k^4-2 k^2
\Omega -8 \Omega ^2\right)\\
& +4 g_2^2\left(k^2+2 \Omega \right) \left(3 k^2+2
\Omega \right) \left(k^2-2 \Omega \right) \\
&+2 g_1^2 g_2
\left(6 k^4+8 k^2 \Omega \right)\\
& +2 g_1 g_2\left(k^2+2 \Omega \right) \left(7
k^4-4 k^2 \Omega -4 \Omega ^2\right)\Big] \; ,
\end{split}
\end{equation}
where
\begin{equation}
\begin{split}
F(k)= & -2 g_2 \left(k^2+2 \Omega \right)+2 g_1 k^2\\
& +\left(3
k^2+2 \Omega \right) \left(k^2-2 \Omega \right)\;.
\end{split}
\end{equation}

Following the same method as in Section
\ref{sec:order2}, we write
\begin{equation}
\label{eq:ansatz_3}
\begin{split}
\Xi^{(3)}=&\;\;  \overline\Xi_0^{(3)}(X,T_1,T_2) +\\
& \left[\widetilde\Xi^{(3)}_1(X,T_1,T_2)  \e^{i \beta(x,t)}+\cc \right] +\\
& \left[\widetilde\Xi^{(3)}_2(X,T_1,T_2)  \e^{2 i \beta(x,t)}+ \cc \right]+\\
& \left[\widetilde\Xi^{(3)}_3(X,T_1,T_2)  \e^{3 i \beta(x,t)}+ \cc \right]
\; .
\end{split}
\end{equation}
We will not need to consider the contribution of the second and third
harmonics in \eqref{eq:order3} and \eqref{eq:ansatz_3}. But the
contributions of the first harmonic ($\widetilde{D}_1(X,T_1,T_2)$ and
$\widetilde\Xi^{(3)}_1$) and of the ``zero''-harmonic
($\overline{D}_0(X,T_1,T_2)$ and $\overline\Xi^{(3)}_0$) are
important. Reinserting expression \eqref{eq:ansatz_3} in
\eqref{eq:order3} yields:
\begin{equation}
\label{eq:order3_0}
\mathbb{M}_0 \; \overline\Xi_0^{(3)}(X,T_1,T_2)  = \overline{D}_0(X,T_1,T_2)\; ,
\end{equation}
and
\begin{equation}\label{eq:order3_1}
\mathbb{M}_1 \; \widetilde\Xi^{(3)}_1(X,T_1,T_2) = \widetilde{D}_1(X,T_1,T_2)\; .
\end{equation}
Again, for solving Eq.~\eqref{eq:order3_0} one must make sure that
$\overline{D}_0$ is in the image space of $\mathbb{M}_0$: this yields
\begin{equation}
\label{eq:Phi_rho}
\overline{L}^{\, t}_0 \cdot \overline{D}_0 (X,T_1,T_2) = 0\; ,
\end{equation}
which writes
\begin{equation}
\label{eq:Phi_rho_b}
\partial_{T_1}^2 \overline\Phi^{(1)} - c^2 \, \partial_{X}^2
\overline\Phi^{(1)} = S(k) \,
\partial_X \left|\widetilde\theta^{(1)}\right|^2 \; ,
\end{equation}
where $c$ is the speed of sound [cf. Eq. \eqref{eq:sound_v}] and
\begin{equation}\label{Sdek}
S(k)=\frac{(g_1+g_2)(k^2+2\,\Omega+g_2) +
g_1 g_2}{2\,\omega_+(k)/k}\; .
\end{equation}
The solution of \eqref{eq:Phi_rho_b} reads (computations are explained
in Appendix \ref{app1}):
\begin{equation}
\label{sol:Phi_rho}
\overline\Phi^{(1)}(X,T_1,T_2) =W(k) \int^X \!\!\! \mathrm{d}X
\left|\widetilde\theta^{(1)}\right|^2\; ,
\end{equation}
where
\begin{equation}
\label{sol:Phi_rho_b}
W(k)=\frac{S(k)}{[\omega'_+(k)]^2-c^2}\; .
\end{equation}
Expression \eqref{sol:Phi_rho} combined with Eq.~\eqref{eq-importante}
shows that
\begin{equation}
\partial_{T_1} \overline\Phi^{(1)}
+ \omega_+^\prime(k)\, \partial_X \overline\Phi^{(1)} = 0\; .
\end{equation}
This result shows that the deformation of the background propagates
with the group velocity, as does the envelope of the wave [which obeys
the same equation, cf. \eqref{eq-importante}].

Finally, for being able to solve Eq.~\eqref{eq:order3_1} we need
$\widetilde{D}_1$ be in the image space of
$\mathbb{M}_1$:
\begin{equation}\label{eq:GP1_a}
\widetilde{L}^{\, t}_1 \cdot \widetilde{D}_1 (X,T_1,T_2) = 0 \; .
\end{equation}
This reads
\begin{equation}\label{eq:GP1}
\begin{split}
&i\, \partial_{T_2}\, \widetilde\theta^{(1)}  =
-\frac{\omega_+^{\prime\prime}(k)}{2}\, \partial_X^2
\widetilde\theta^{(1)} \\
&+
\Bigg[
\left( P(k)-\frac{Q(k)}{\Delta}
\right) \left|\widetilde\theta^{(1)}\right|^2 \\
& + k
\left(1+\frac{g_2}{g_1}\frac{k^2+2\,\Omega+g_2}{k^2+2\,\Omega+2\,g_2}
\right) \partial_X
\overline\Phi^{(1)} \Bigg] \frac{\widetilde\theta^{(1)}}{2}\; ,
\end{split}
\end{equation}
where $\omega_+^{\prime\prime}(k)={\rm d}^2\omega_+/{\rm d}k^2$.
One can re-express the term
$\partial_X \overline\Phi^{(1)}$ using Eq.~\eqref{sol:Phi_rho}. One
then obtains a nonlinear Schr\"odinger equation (NLS) for
$\widetilde\theta^{(1)}(X,T_1,T_2)$:
\begin{equation}\label{eq:GP2}
i\,\partial_{T_2}\, \widetilde\theta^{(1)} =-
\frac{\omega_+^{\prime\prime}(k)}{2}\, \partial_X^2
\widetilde\theta^{(1)} + g_{\rm eff}(k)
\left|\widetilde\theta^{(1)}\right|^2 \widetilde\theta^{(1)} \;,
\end{equation}
with
\begin{equation}\label{eq:GP3}
\begin{split}
g_{\rm eff}(k) =
& \frac{1}{2} \left( P(k) - \frac{Q(k)}{\Delta} \right)
\\
+ & \frac{k}{2}
\left(1+\frac{g_2}{g_1}\frac{k^2+2\,\Omega+g_2}{k^2+2\,\Omega+2\,g_2}\right)
W(k) \;.
\end{split}
\end{equation}
One has reached a point where the approach is self-contained, as far
as the first order term $\Xi^{(1)}$ in expansion \eqref{ansatz_upper}
is concerned. One just needs to return to the actual variables $x$ and
$t$ using the reverse of transformations \eqref{derivatives}
\cite{rembackxt}.  We give below final formulas valid even when
$k_0\neq 0$.

\subsection{Final formulas and discussion}\label{final-upper}

A nonlinear wave packet propagating in the upper branch is described
by a set of fields $\Xi(x,t)$ of the form \eqref{mod7bis} with
\begin{equation}\label{eq:GP4}
\Xi'(x,t)=\overline\Xi(x,t) +
\left[\widetilde\Xi(x,t)\e^{i(k x -\omega_+(k)t)} + \cc\right] \; .
\end{equation}
The component $\widetilde{\theta}(x,t)$ of the envelope
$\widetilde\Xi(x,t)$ is solution of
\begin{equation}\label{eq:GP5}
i\,\partial_{t}\,\widetilde\theta =-
\frac{\omega_+^{\prime\prime}(k)}{2}\, \partial_y^2
\,\widetilde\theta + g_{\rm eff}(k)
\left|\,\widetilde\theta\,\right|^2 \widetilde\theta \;.
\end{equation}
where $y=x-\omega'_+(k) t$ is the space coordinate in a frame moving
at the group velocity.

Once $\widetilde\theta(x,t)$ has been determined,
the component $\overline\Phi(x,t)$ of the background deformation
is obtained as
\begin{equation}\label{eq:GP6}
\overline\Phi(x,t)=W(k) \int^x\!\! \mathrm{d}x
  \left|\,\widetilde\theta\,\right|^2\; .
\end{equation}
The other components of the background and of the envelope are given by
\begin{equation}\label{eq:GP7}
\overline\Xi(x,t) =
\begin{pmatrix}
\overline{n}(x,t)\\ \overline\Phi(x,t) \\
\overline\theta(x,t)\\ \overline\varphi(x,t)
\end{pmatrix}
= \overline{R}\; \overline\Phi(x,t)
\; ,
\end{equation}
and
\begin{equation}\label{eq:GP8}
\widetilde\Xi(x,t)=
\begin{pmatrix}
\widetilde{n}(x,t)\\ \widetilde\Phi(x,t)\\
\widetilde\theta(x,t)\\ \widetilde\varphi(x,t)
\end{pmatrix}
=  \widetilde{R} \; \widetilde\theta(x,t) \; ,
\end{equation}
where $\overline{R}^t=(0,1,0,0)$ and
\begin{equation}\label{eq:GP9}
\widetilde{R}=\begin{pmatrix}
\frac{k \left(k^2+2 \Omega \right) \left(2 g_2+k^2-4 k_0^2+2
\Omega \right)-4 k\,\omega_+^2(k)}{8 k_0 \left(k^2+\Omega \right)\omega_+(k)}
\\[1mm]
-\frac{i \left((2 g_2+k^2+2 \Omega) (k^2+2 \Omega )+4 k^2
k_0^2-4 \omega_+(k) ^2\right)}{4 k k_0 \left(k^2+\Omega \right)} \\[1mm]
1\\[1mm]
\frac{i \left(k^2 (2 g_2+k^2-4 k_0^2+2 \Omega )+4
\omega_+^2(k)\right)}{4  \left(k^2+\Omega \right)\omega_+(k)}
\end{pmatrix}\; .
\end{equation}
We do not write here the explicit forms of $W(k)$ and $g_{\rm eff}(k)$
for $k_0\ne 0$ because they are too cumbersome. However, it is
important for subsequent discussions
to stress that Eq.~\eqref{sol:Phi_rho_b} still holds for
$k_0\neq 0$, but with a numerator $S(k)$ whose expression is different
from the one given in Eq. \eqref{Sdek} for the case $k_0=0$. On the
other hand, the formulas \eqref{eq:GP7}, \eqref{eq:GP8}, and
\eqref{eq:GP9} are valid even when $k_0\neq 0$. Note that
$\overline{R}$ is identical to $\overline{R}_0$ defined in
\eqref{order1c} and that $\widetilde{R}$ reduces to $\widetilde{R}_1$
defined in \eqref{order1d} when $k_0=0$. In this case, the first two
components of $\widetilde{R}$ cancel and the
nonlinear structure corresponds to a polarization
signal, with oscillations of $\rho_\uparrow$ and $\rho_\downarrow$
preserving a fixed total density.

The nonlinear Schr\"odinger equation \eqref{eq:GP5} describes the
spatio-temporal evolution of the envelope wave which is advected by
the group velocity $\omega'_+(k)$ while dispersion and nonlinearity
give corrections to the dynamics of the wave train, in particular for
large times. It has been obtained through a multiscale expansion
assuming the existence of well separated spatial and temporal
scales. The two spatial scales are the wave length $\sim k^{-1}$
(associated with the $x$-dependence of the phase $\beta(x,t)$
\eqref{beta}) and the length $a$ characteristic of the spatial
variations of the envelope of the wave packet (associated to
coordinate $X$). These length scales
should be widely different and this corresponds
to defining our small parameter as
\begin{equation}\label{ak}
\epsilon=\frac{1}{a k} \ll 1 \; .
\end{equation}
The three time scales legitimating the introduction of the three
different time coordinates $t$, $T_1$ and $T_2$ are :
\begin{equation}\label{GP10}
\tau \ll
\tau_1 \ll
\tau_2  \; ,
\end{equation}
where $\tau\sim 1/\omega_+(k)$ is the period of the carrier wave. The
characteristic time $\tau_1$ is associated to the group motion of the
envelope.  The time $\tau_2$ accounts for the fact that the envelope
not only propagates with the group velocity, but also changes form
because of higher order dispersive effects and of nonlinearity (both
effects typical balance in a nonlinear wave such as the soliton
solutions discussed below). From Eqs. \eqref{eq-importante} and
\eqref{eq:GP2} one can check that when the condition \eqref{ak} is
fulfilled one has $\tau/\tau_1\sim \tau_1/\tau_2\sim \epsilon$, thus
legitimating {\sl a posteriori} the introduction of the three time
coordinates \eqref{multiscale0}.

When $g_{\rm eff}(k)$ is positive, periodic wave trains with constant
amplitude formed in the upper branch of the spectrum are dynamically
stable. They can support nonlinear excitations such as dark
solitons. In this case $\widetilde{\theta}$ -- solution of
\eqref{eq:GP5} -- is of the form
\begin{equation}\label{dark}
\widetilde{\theta}(y,t)=\Theta_{0} \, \e^{-i\, g_{\rm eff}(k)\, \Theta_{0}^2\, t}
\left[\cos\alpha\tanh(Y)+i\sin\alpha\right]\; ,
\end{equation}
where $\Theta_{0}\in\mathbb{R}^+$ is the amplitude of the wave
train and $\alpha\in[0,\pi/2]$; $\sin\alpha$ is the dimensionless
velocity of the dark soliton, cf. Eq.~\eqref{argument}. The argument
$Y$ in \eqref{dark} is
\begin{equation}\label{argument}
Y=\frac{y-V_{\rm sol}\, t}{\xi_{\rm eff}(k)}\cos\alpha\; ,  \; \mbox{where}\;
V_{\rm sol}=\sin\alpha\, c_{\rm eff}(k) \; ,
\end{equation}
and $c_{\rm eff}(k)=\Theta_{0}\sqrt{g_{\rm
    eff}(k)\omega_+''(k)}=\omega_+''(k)/ \xi_{\rm eff}(k)$.

\begin{figure}
\includegraphics[width=0.99\linewidth]{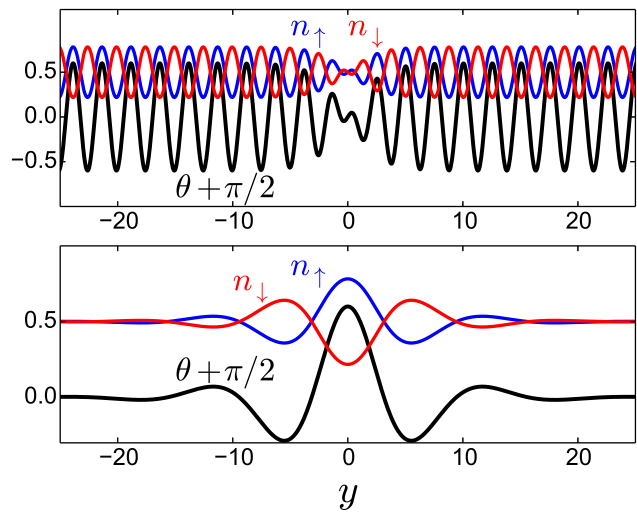}
\caption{(Color online). $\theta$, $n_\uparrow$ and $n_\downarrow$ as
  functions of $y$ for a dark envelope soliton \eqref{dark} (upper
  plot) and a bright envelope one \eqref{bright} (lower plot). The
  system's parameters are the same as in Fig.~\ref{fig-geff1}.  For
  both plots $\Theta_{0}=0.3$. The dark soliton is plotted for
  $k=2.5$  ($g_{\rm eff}(k)=1.4$)
and $V_{\rm sol}=0$ (black soliton) and the bright soliton for
  $k=0.5$ ($g_{\rm eff}(k)=-0.6$).}\label{fig-soliton}
\end{figure}

In the case where $g_{\rm eff}(k)<0$, wave trains in the upper branch
are dynamically unstable (they experience a modulational instability,
see below), but one may observe stable bright envelope solitons, for
which the solution of \eqref{eq:GP5} is of the form
\begin{equation}\label{bright}
\widetilde{\theta}(y,t)=
\frac{\Theta_{0} \exp(-i\, \frac{g_{\rm eff}(k)}{2}\, \Theta_{0}^2\, t)}
{\cosh\left(\Theta_{0} \sqrt{\frac{-g_{\rm eff}(k)}{\omega_+''(k)}}
\, y\right)}
\; ,
\end{equation}
where $\Theta_{0}$ is a positive real parameter governing the
amplitude of the soliton.  Once $\widetilde{\theta}$ is known, the
corresponding values of the other fields describing the system are
then given by \eqref{eq:GP6}, \eqref{eq:GP7} and \eqref{eq:GP8}. The
density profiles of typical envelope solitons are plotted in
Fig. \ref{fig-soliton}.

\begin{figure}[h]
\includegraphics[width=0.99\linewidth]{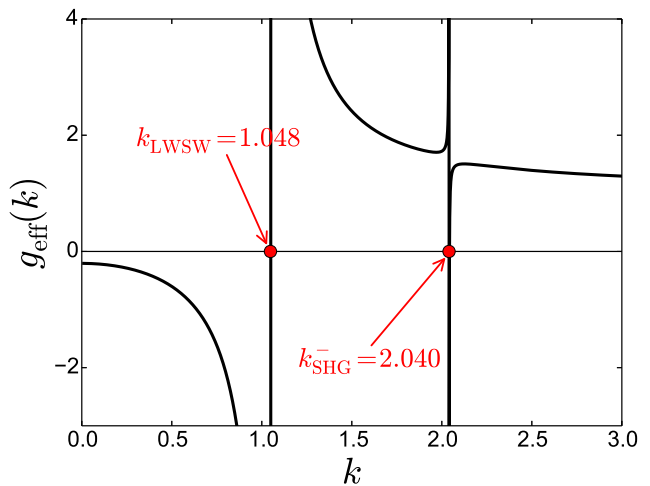}
\caption{(Color online). The solid line represents $g_{\rm eff}(k)$
  for the choice of parameters $k_0=0$, $g_1=2.2$, $g_2=0.2$ and
  $\Omega=2.5$.}\label{fig-geff1}
\end{figure}

In order to get better insight on the type of dynamics described by
the envelope NLS equation \eqref{eq:GP5}, we show in
Fig.~\ref{fig-geff1} how the effective nonlinear constant $g_{\rm
  eff}$ depends on $k$.
We see that $g_{\rm eff}(k)$ starts at low $k$ with a negative value,
and since $\omega^{\,\prime\prime}_+(k)>0$, this means that wave
trains in the upper branch experience a modulational instability, see,
e.g., Ref.~\cite{Zak09} and references therein. Modulational
instability in Bose-Einstein condensates with repulsive interaction
have already been studied in the presence of an external optical
lattice potential \cite{mi-ol} and for the counterflow of two miscible
species \cite{mi-ab}. Here we consider a scenario closer to the
original Bejamin-Feir configuration \cite{Benj-Feir}, where
nonlinearity destabilizes a periodic wave-train through generation of
spectral sidebands [see the discussion below, around Eqs.~\eqref{mi1}
and \eqref{mi2}].

When discussing the physical origin of the modulational instability in
the present context, it is interesting to note that $g_{\rm eff}(k)$
diverges and changes sign for a value of $k$ which is denoted as
$k_{\rm LWSW}$ in Fig.~\ref{fig-geff1}. For this value of $k$, the system
displays a so called long wave-short wave resonance \cite{LWSW}. In
the present configuration this corresponds to a case where the wave in
the upper branch (with wave vector $k$) decays into two waves,
one in the same branch with a similar wave vector ($k'$), and an other
one in the lower branch, with a small wave vector ($q$, the ``long
wave''). The conditions of conservation of momentum and energy read
$k=k'+q$ and
\begin{equation}\label{lwsw1}
\omega_+(k)=\omega_+(k-q)+\omega_-(q) \; .
\end{equation}
Since $q$ is small one can expand the first term of the r.h.s. of
\eqref{lwsw1} as: $\omega_+(k-q) \simeq \omega_+(k)-q\,\omega_+'(k)$, and
also write $\omega_-(q)\simeq c \, q$. Hence, the phenomenon occurs at
$k=k_{\rm LWSW}$ such that
\begin{equation}\label{lwsw2}
\omega'_+(k_{\rm LWSW})=c \; ,
\end{equation}
meaning that the condition of resonance is that the group velocity of
the short wave is equal to the phase velocity of the long wave.

The location of the resonance is clearly seen in Fig.~\ref{fig-geff1},
at a value of $k$ in exact agreement with the value $k_{\rm LWSW}$
determined by \eqref{lwsw2}. From the derivation leading to the NLS
Eq.~\eqref{eq:GP5}, one can locate the mathematical origin of the
resonance phenomenon in Eq.~\eqref{eq:GP6}, where $W(k)$ as given by
\eqref{sol:Phi_rho_b} clearly diverges exactly at resonance. The
phenomenological analysis just presented assumes that this divergence
corresponds to a transfer of excitation from the upper branch to the
lower one, but one should ascertain that this is indeed the case in
our mathematical treatment. Indeed, it might seem from
Eq.~\eqref{eq:Phi_rho_b} that the divergence is connected to a
resonance with a deformation of the background (of
$\overline{\Phi}^{(1)}$) which might not be exactly connected to the
lower branch of excitation. A first clue of this connection comes from
the l.h.s. of Eq.~\eqref{eq:Phi_rho_b} itself: in this equation, the
zero mode of the operator acting on $\overline{\Phi}^{(1)}$
corresponds to a dispersion relation which is the long wave length
approximation of the lower branch: $\omega'_-(k)\simeq c\, k$. The
second and final reason explaining why in this context,
$\overline{\Phi}^{(1)}$ indeed represents the lower branch excitation
comes from the very reason for its appearance in \eqref{eq:Phi_rho_b}:
it originates from Eq.~\eqref{order1c}, more precisely, from the
specific form of $\overline{\Xi}^{(1)}_0$ which is tailored to be
representative of the kernel of $\mathbb{M}_0$. And, as can be checked
by a comparison of the forms and definitions of $\mathbb{M}_1$
\eqref{linear2} and $\mathbb{M}_0$ \eqref{eq:M0}, $\mathbb{M}_0$ is
the $k\to 0$ limit of $\mathbb{M}_1$ when $\omega=\omega_-(k)$: hence
the background contributions in the ansatz \eqref{init-ansatz} (and in
the higher order terms) is indeed a low $k$ contribution in the lowest
branch and the divergence of $W(k)$ in \eqref{sol:Phi_rho_b} indeed
corresponds to a resonance between the upper branch and the (long
wavelength limit of) the lower branch.

It is remarkable that the occurrence of the long wave-short wave
resonance is connected to a disappearance of the modulational
instability of the upper branch: as one can see from
Fig.~\ref{fig-geff1} the nonlinear parameter $g_{\rm eff}(k)$ is
positive when $k$ is larger than $k_{\rm LWSW}$ and
wave trains in the upper branch are thus stable when their wave-vector
is larger than the one of the long wave-short wave resonance. In order to
appreciate the origin of this phenomenon one first needs to get some
physical insight on the cause of the modulational instability.  Since
the reasoning presented below is quite general, and for simplifying
the notations, we will here for a moment denote the dispersion
relation as $\omega(k)$ instead of $\omega_+(k)$.

\begin{figure}[h]
\includegraphics[width=0.99\linewidth]{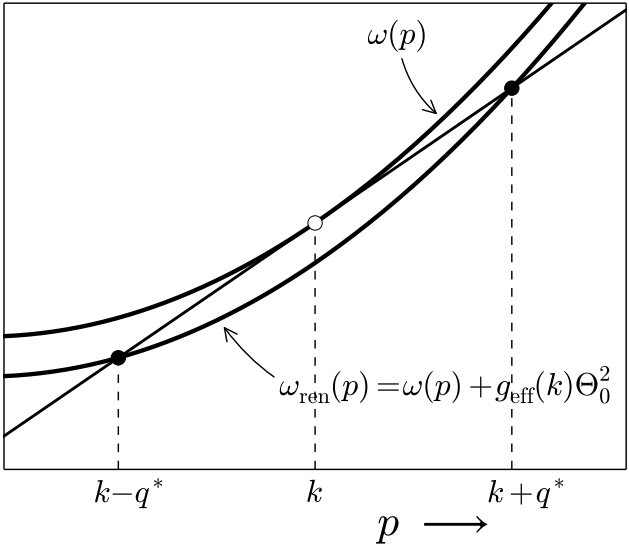}
\caption{Illustration of the modulational instability process. The
  straight solid line is the tangent to $\omega(p)$ at $p=k$. Two
  initial elementary excitations $(k,\omega(k))$ can decay into
  $(k-q^*,\omega_{\rm ren}(k-q^*))$ and $(k+q^*,\omega_{\rm
    ren}(k+q^*))$ provided $q^*$ verifies the above construction,
  i.e., that the two black points are the intersections of the
  straight solid line with the renormalized dispersion relation
  $\omega_{\rm ren}(p)$. An expansion of $\omega(p)$ at second order
  in the vicinity of $k$ shows that (i) the white point is the middle
  of the two black points (and this automatically implies momentum
  and energy conservation in the process \eqref{mi3}) and that (ii)
  $q^*$ defined by the above construction verifies
  Eq.~\eqref{mi2}.}\label{fig-MI}
\end{figure}

If one studies a wave train with wave vector $k$ and constant (real)
amplitude $\Theta_0$, one finds from \eqref{eq:GP5} that the
corresponding $\widetilde\theta(x,t)$ is equal to $\Theta_0 \exp\{-i
g_{\rm eff}(k) \, \Theta_0^2 \, t\}$. Then, a perturbative treatment
of Eq.~\eqref{eq:GP5} readily shows (see, e.g.,
Refs.~\cite{Kiv98,Pet02,Pit03}) that small amplitude modulations of
the carrier wave with relative wave vector $q$ and angular frequency
$\varpi$ obey the dispersion relation
\begin{equation}\label{mi1}
\left(\varpi - \omega'(k) \, q\right)^2 =
\left(\frac{\omega^{\,\prime\prime}(k)\, q^2}{2}\right)^2 +
g_{\rm eff}(k) \, \Theta_0^2 \, \omega^{\,\prime\prime}(k)\, q^2\; .
\end{equation}
If $g_{\rm eff}(k)$ is negative, $\varpi$ will be imaginary (for low
enough values of $q$), meaning that the wave train is dynamically
unstable. The value $q^*$ of $q$ corresponding to the largest
imaginary part of $\varpi$, i.e., to the greatest growth rate of the
perturbations, verifies
\begin{equation}\label{mi2}
\frac{\omega^{\,\prime\prime}(k)}{2}(q^*)^2
=-g_{\rm eff}(k)\,\Theta_0^2 \; .
\end{equation}
One gets here a confirmation that the wavetrain is unstable when
$\omega^{\,\prime\prime}(k)\,g_{\rm eff}(k)$ is negative. This
corresponds to the so-called Lighthill-Benjamin-Feir criterion of
modulational instability \cite{Zak09}, which can be given the
following intuitive interpretation: one assumes that a wave-train of
finite amplitude $\Theta_0$ corresponds to the renormalized
dispersion relation
\begin{equation}
\omega_{\rm ren}(k)=\omega(k)+g_{\rm eff}(k)\,
\Theta_0^2 \; .
\end{equation}
The initial carrier wave at angular frequency $\omega(k)$
and wave-vector $k$ may decay into two side bands according to the
following process:
\begin{equation}\label{mi3}
\begin{split}
\omega(k) + \omega(k) & \to \omega_{\rm ren}(k-q^*)+
\omega_{\rm ren}(k+q^*)\; , \\
k + k & \to (k-q^*) + (k+q^*) \; ,
\end{split}
\end{equation}
where $q^*$ as given by \eqref{mi2} enforces the energy and momentum
conservation relations in the process \eqref{mi3}, as can be checked
analytically (by an expansion in $q^*$) and is graphically demonstrated
in Fig.~\ref{fig-MI}. It is clear that, when $\omega''(k)>0$, the
geometrical construction of Fig.~\ref{fig-MI} is only possible if
$g_{\rm eff}(k)<0$: in this case the wave train is modulationally
unstable.

Hence we understand why the change of sign of $g_{\rm eff}(k)$
observed in Fig.~\ref{fig-geff1} when $k$ crosses $k_{\rm LWSW}$
changes the stability of the wave train. Now it remains to understand
the physical reason for this change of sign. Actually, the reason for
it becomes clear when one focuses on the nonlinear term in brackets in
Eq.~\eqref{eq:GP1}. The first part of this term [with the $P(k)$ and
$Q(k)$ contributions] is a genuine nonlinear self-interaction, but the
second part is proportional to $k \,\partial_X\overline{\Phi}^{(1)}$,
i.e., to $k \, \overline{U}^{(1)}$ which is a Doppler contribution to
the energy of an excitation moving over a background of velocity
$\overline{U}^{(1)}$. For $k<k_{\rm LWSW}$, the momentum $q$ imparted
to the lower branch is negative, and the corresponding value of
$\overline{U}^{(1)}$ is also, as physically clear and mathematically
demonstrated by the fact that in this case $W(k)<0$ [see
Eq.~\eqref{sol:Phi_rho_b}]. It so happens that this Doppler
contribution is dominant over the self-interaction terms, and, as a
result, $g_{\rm eff}(k<k_{\rm LWSW})<0$. On the contrary, for
$k>k_{\rm LWSW}$ the momentum imparted to the lower branch is
positive, $\overline{U}^{(1)}>0$ and $g_{\rm eff}(k>k_{\rm
  LWSW})>0$. This ends our discussion of the behavior of $g_{\rm
  eff}(k)$ around $k\simeq k_{\rm LWSW}$ and the explanation for the
disappearance of the modulational instability when $k \gtrsim k_{\rm
  LWSW}$.

Besides the long wave-short wave resonance, one can notice an other
resonant-like structure in Fig.~\ref{fig-geff1}. It corresponds to a
generation of second harmonic according to the three waves process
\begin{equation}\label{shg1}
\begin{split}
k + k & \to 2k \; , \\
\omega_+(k)+\omega_+(k)& \to \omega_-(2k) \; .
\end{split}
\end{equation}
The condition of conservation of momentum and energy in the above
process determines the value of the resonant wave vector
$k_{\rm SHG}^-$ in excellent agreement with the location of the divergence
of $g_{\rm eff}(k)$ observed in Fig.~\ref{fig-geff1}. In the vicinity
of $k_{\rm SHG}^-$ our approach fails (and the envelope NLS equation
\eqref{eq:GP5} is not relevant) because the determinant of
$\mathbb{M}_2$ vanishes, contrarily to what has been stated after
Eq.~\eqref{def:M2}, and the procedure that has been used for
determining $\widetilde\Xi^{(2)}_2$ from Eq.~\eqref{eq:order2_2} is
incorrect. In this case the assumption that higher-order
harmonics have a very small contribution is wrong.
The fact that second harmonic generation
is associated with vanishing of the determinant of $\mathbb{M}_2$ is
an immediate result of the definition \eqref{def:M2} and of energy
conservation in the process \eqref{shg1}: at resonance one has
$\mathbb{M}_2\equiv \mathbb{M}(2\,i\,k_{\rm SHG}^-, -2 i\,\omega_+(k_{\rm SHG}^-) )
= \mathbb{M}(2\,i\,k_{\rm SHG}^-, - i \omega_-\,(2 k_{\rm SHG}^-) ) $. The
determinant of this last matrix is zero, because, for any $p$, $\det
\mathbb{M}(i p, - i \omega_- (p) ) =0$, since $\omega=\omega_-(p)$
is one of the dispersion relations of the system.

\begin{figure}[h]
\includegraphics[width=0.99\linewidth]{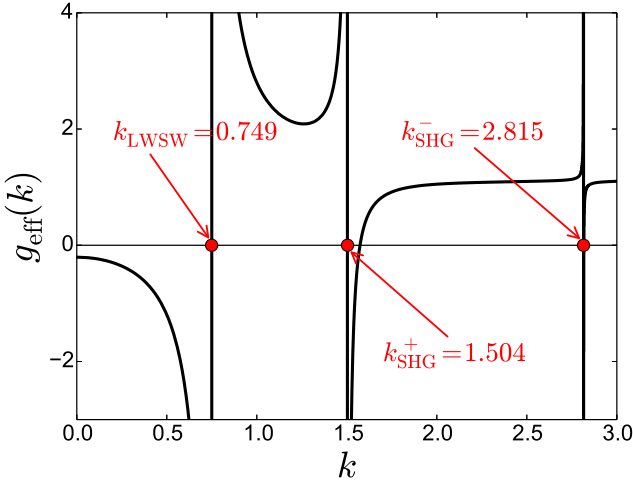}
\caption{(Color online). The solid line represents $g_{\rm eff}(k)$
  for the choice of parameters $k_0=0.5$, $g_1=2.2$, $g_2=0.2$ and
  $\Omega=2.5$. The location of the resonances is determined by
  momentum and energy conservation in the processes \eqref{lwsw1} (for
  $k_{\rm LWSW}$), \eqref{shg1} (for $k_{\rm SHG}^-$) and \eqref{shg2}
  (for $k_{\rm SHG}^+$).}\label{fig-geff2}
\end{figure}

For concluding the discussion, it is interesting to notice that,
besides the second harmonic generation identified in
Fig.~\ref{fig-geff1}, there exists an other possible generation of
second harmonics, which only involves excitations of the upper branch:
\begin{equation}\label{shg2}
\begin{split}
k + k & \to 2k \; , \\
\omega_+(k)+\omega_+(k)& \to \omega_+(2k) \; .
\end{split}
\end{equation}
This new process should induce a divergence of $g_{\rm eff}(k)$ at the
wave vector $k=k_{\rm SHG}^+$ which ensures energy conservation in the
process \eqref{shg2}. Indeed, in this case we have a linear system
$\mathbb{M}(2\,i\,k_{\rm SHG}^+, -2 i\,\omega_+(k_{\rm SHG}^+) ) =
\mathbb{M}(2\,i\,k_{\rm SHG}^+, - i \omega_+\,(2 k_{\rm SHG}^+) ) $
that has a zero determinant because $\omega=\omega_+(p)$ is one of the
dispersion relations of the system.  For the set of parameters
corresponding to Fig.~\ref{fig-geff1}, this second harmonic generation
should occur at $k_{\rm SHG}^+= 1.612$. It is then surprising that
this resonance is not seen in this figure.  However, it is clearly
seen when $k_0\neq 0$ (see Fig.~\ref{fig-geff2}), at the value
predicted by the conservation of energy in \eqref{shg2}.

Actually, in the case where $k_0=0$, at $k=k_{\rm SHG}^+$ one has
$\det\mathbb{M}_2=0$, and the divergent factor involved in the
determination of $\widetilde\Xi^{(2)}_2$ from Eq.~\eqref{eq:order2_2}
(and which results in a divergence in the expression of
$g_{\rm eff}(k)$) is canceled by an other contribution. This can be easily
understood by noticing that $\mathbb{M}$ defined in Eq.~\eqref{mod9}
is a block matrix when $k_0=0$:
\begin{equation}
\mathbb{M}=
\begin{pmatrix}
\mathbb{M}_- & 0 \\
0 & \mathbb{M}_+
\end{pmatrix}
\end{equation}
where $\mathbb{M}_-$ and $\mathbb{M}_+$ are $2\times 2$ matrices
accounting for the lower and the upper excitation branches.
We are here interested in second harmonic generation, i.e., in the
specific matrix $\mathbb{M}_2=\mathbb{M}(2 i k, -2 i \omega_+(k))$. In
this case, we denote the matrices $\mathbb{M}_-$ and $\mathbb{M}_+$ as
$\mathbb{M}_{2-}$ and $\mathbb{M}_{2+}$ and their inverses are
\begin{subequations}
\begin{align}
&\mathbb{M}_{2-}^{-1} =
\frac{1}{\omega_-^2(2k)-(2\omega_+(k))^2}
\, \mathrm{adj}\left(\mathbb{M}_{2-} \right)\; , \label{m2--}\\
&\mathbb{M}_{2+}^{-1} =
\frac{1}{\omega_+^2(2k)-(2\omega_+(k))^2}
\,\mathrm{adj}\left(\mathbb{M}_{2+} \right)\; ,\label{m2-+}
\end{align}
\end{subequations}
where ``adj'' denotes the adjugate matrix. The divergence of
$g_{\rm eff}(k_{\rm SHG}^+)$ is associated with the divergence of the
denominator in \eqref{m2-+}, corresponding to energy conservation in
the process \eqref{shg2}. In the special case $k_0=0$, the solution of
Eq. \eqref{eq:order2_2} reads:
\begin{equation}\label{shg2+1}
\widetilde\Xi^{(2)}_2 =
\begin{pmatrix}
\mathbb{M}_{2-}^{-1} & 0 \\
0 & \mathbb{M}_{2+}^{-1}
\end{pmatrix}
\, \widetilde{C}_2\; ,
\end{equation}
where $\widetilde{C}_2$ is given by Eq.~\eqref{lesc2} when $k_0=0$: in
this case its last two components are zero. Eq.~\eqref{shg2+1} then reads
\begin{equation}
\widetilde\Xi^{(2)}_2=
\begin{pmatrix*}[l]
\mathbb{M}_{2-}^{-1}
\begin{pmatrix}
i\,\Delta\, k^2  \\ \frac{1}{2} \left(
\frac{k^2-2\,\Omega}{k^2+2\,\Omega}\,g_2 - k^2 - 2\,\Omega \right)
\end{pmatrix}\\
\mathbb{M}_{2+}^{-1}
\begin{pmatrix}
0 \\ 0
\end{pmatrix}=\begin{pmatrix}
0 \\ 0
\end{pmatrix}
\end{pmatrix*}\left(\widetilde\theta^{(1)}\right)^2 \; ,
\end{equation}
and the possible divergence of the denominator of
$\mathbb{M}_{2+}^{-1}$ is masked. This is the reason for the
inhibition of the second harmonic generation process \eqref{shg2} when
$k_0=0$.

\section{Nonlinear perturbation theory for excitations
propagating  in the lower branch}\label{sec:lowerbranch}

\begin{figure}[h]
\includegraphics[width=0.99\linewidth]{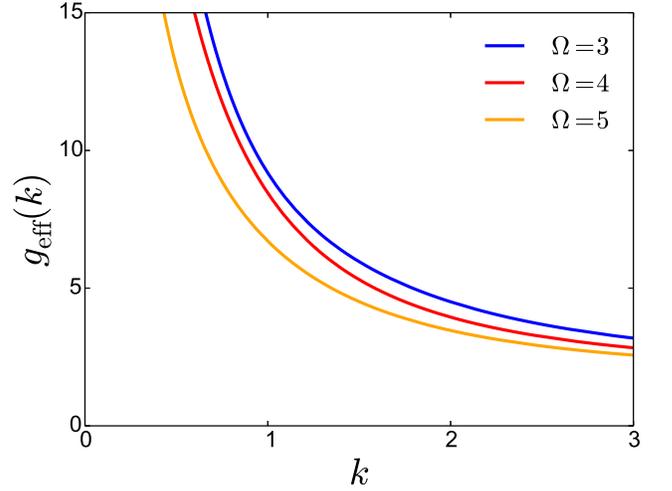}
\caption{(Color online). Nonlinear coefficient $g_{\rm eff}(k)$ for
  the envelope NLS equation describing a wave packet propagating in
  the lower excitation branch. The curves are drawn for different
  values of $\Omega$; the other parameters are $k_0=1$, $g_1=2.2$
  and $g_2=0.2$.}\label{fig-geff-LB}
\end{figure}

We now study the propagation of a sound pulse which, in a linear
approximation, would lie on the lower excitation branch. The method
used in section \ref{sec:upperbranch} can be employed in the present
case. It yields for the nonlinear coefficient
$g_{\rm eff}(k)$ a behavior represented in
Fig. \ref{fig-geff-LB}.
The nonlinear coefficient diverges at large wave-length. This is due
to the fact that, for the lower branch, the analog of the coefficient
$W(k)$ \eqref{sol:Phi_rho_b} diverges when $k\to 0$ since
$\omega_-'(0)=c$.  In this case the nonlinear time $t_{\rm NL}\propto
g_{\rm eff}(k)^{-1}$ associated to Eq. \eqref{eq:GP5} diverges
indicating that nonlinear structures form extremely rapidly.  $t_{\rm
  NL}$ may become even smaller that the period of the wave (except for
waves of extremely small amplitude) and in this case the technique of
the envelope NLS fails.

In this long wave length limit one can suggest an alternative method
consisting in deriving equations for the interacting fields themselves
instead of an effective equation for the envelope.
This method is based on
the following reasoning: In the linear regime and at the level of
accuracy at which the expansion \eqref{sol:disp} holds, any of the
components of $\Xi'(x,t)$ -- $n'$ say -- satisfies the linear equation
\begin{equation}\label{eq17}
    n'_t+c\, n'_x-c_3\,n'_{xxx}=0\; ,
\end{equation}
where the last term describes a small dispersive correction to the
propagation with constant velocity $c$.  If the amplitude $n'$ is
small but finite and such that this term has the same order of
magnitude as the leading nonlinear correction to \eqref{eq17} (which
is typically quadratic in $n'$), then nonlinear effects cannot be
omitted for correctly describing the propagation of the pulse. In this
regime one can try to derive an equation of the type \eqref{eq17} with
additional terms taking into account weak nonlinear effects. The most
natural extension of \eqref{eq17} is a Korteweg-de Vries (KdV) equation
in which a nonlinear term of the form $n' n'_x$ accounts for a
dependence in density fluctuations ($\propto n'$) of the velocity of
sound.

\subsection{Quadratic nonlinearity: KdV regime}\label{KdVregime}

It now is appropriate to work in a reference frame moving at the
speed of sound $c$, and to use $x-ct$ and $t$ as coordinates. In
order that the derivatives in equations of type \eqref{eq17} appear at the
same order, we define
\begin{equation}\label{kdv1}
\xi=\epsilon^{1/2}(x-ct) \; , \quad\mbox{and}\quad \tau=\epsilon^{3/2} t \; .
\end{equation}
where $\epsilon$ will henceforth be a small positive parameter. The
choice of the specific powers $\epsilon^{1/2}$ and $\epsilon^{3/2}$ in
\eqref{kdv1} (instead of $\epsilon$ and $\epsilon^{3}$ for instance)
will make sure that the derivatives in equations of type \eqref{eq17}
appear at the same order as the quadratic nonlinear contribution
[$\propto n'n'_x$, see Eq.~\eqref{kdv14} below]. In terms of the new
variables $\xi$ and $\tau$ and of the velocities $U$ and $v$ defined
in Eq.~\eqref{kdv0}, the system \eqref{mod5} reads
\begin{subequations}\label{kdv2}
\begin{align}
\epsilon^{3/2}n_\tau=\;& \epsilon^{1/2}c\,n_\xi +
\epsilon^{1/2} k_0 \left[n \cos\theta\right]_\xi \nonumber  \\
-\;&
\tfrac12 \epsilon^{1/2} \big[n\left(U-v\cos\theta \right)\big]_\xi \;,
 \\
\epsilon^{3/2}U_\tau  =\; &
\epsilon^{1/2}c\,U_\xi - \epsilon^{1/2} k_0 \, v_\xi \nonumber \\
-\;& \epsilon^{1/2}
 \left[\frac{\epsilon}{4} \frac{n_\xi^2}{n^2}
- \frac{\epsilon}{2} \frac{n_{\xi\xi}}{n} - \epsilon \frac{\cot \theta}{2}
\frac{\left[n \, \theta_\xi \right]_\xi}{n}\right]_\xi \nonumber \\
	-\;&\epsilon^{1/2}
\left[\frac{\epsilon\theta_\xi^2+U^2+v^2}{4}+ g_1 n \right]_\xi
 \nonumber  \\
+\;&\epsilon^{1/2}\theta_\xi\frac{\Omega \cos\theta}{\sin^2\theta}
\cos\varphi +\frac{\Omega}{\sin\theta} \, v\, \sin\varphi \;,
 \\
\epsilon^{3/2}\theta_\tau  =\; & \epsilon^{1/2}c \, \theta_\xi
- \Omega \, \sin\varphi \nonumber \\
-\;& \tfrac12 \epsilon^{1/2} \left(U \theta_\xi
+ \frac{\left[n\,(v+2k_0)\sin\theta \right]_\xi}{n} \right) \;,
\\
\epsilon^{3/2}v_\tau=\; & \epsilon^{1/2}c\,v_\xi - \epsilon^{1/2} k_0 \, U_\xi
 \nonumber  \\
+\;& \epsilon^{1/2}
\left[\frac{\epsilon}{2 \sin \theta}
\frac{\left[n \, \theta_\xi \right]_\xi}{n}-\frac{U\,v}{2}
+ g_2 n
\cos\theta\right]_\xi \nonumber  \\
	+\;&\epsilon^{1/2}\theta_\xi\,
\frac{\Omega \cos\varphi}{\sin^2\theta}
+ \Omega \, v\, \cot\theta\, \sin\varphi \;.
\end{align}
\end{subequations}
We perform a multi-scale analysis by expanding $(n,U,\theta,v)$ in the
following way:
\begin{equation}\label{kdv3}
\begin{pmatrix}
n(\xi,\tau)\\U(\xi,\tau)\\\theta(\xi,\tau)\\v(\xi,\tau)
\end{pmatrix}
=
\begin{pmatrix}
1\\0\\-\frac{\pi}{2}\\0
\end{pmatrix}
+\epsilon
\begin{pmatrix}
n^{(1)}\\U^{(1)}\\\theta^{(1)}\\v^{(1)}
\end{pmatrix}
+\epsilon^2
\begin{pmatrix}
n^{(2)}\\U^{(2)}\\\theta^{(2)}\\v^{(2)}
\end{pmatrix}
 + \cdots
\end{equation}
In agreement with the definitions \eqref{kdv0} and \eqref{kdv1} we have
\begin{equation}
U=\epsilon^{1/2}\Phi_\xi\; ,\quad\mbox{and}\quad v=\epsilon^{1/2}\varphi_\xi\; .
\end{equation}
Taking into account expansion \eqref{kdv3} and the leading order
\eqref{mod7} for $\Phi$ and $\varphi$ we thus have
\begin{equation}\label{kdv4}
\begin{array}{ccr}
\Phi=& -2\epsilon^{-3/2}\mu\tau & + \epsilon^{1/2}\Phi^{(1)}
+\epsilon^{3/2}\Phi^{(2)} + \cdots\\
\varphi=&  & \epsilon^{1/2}\varphi^{(1)}
+\epsilon^{3/2}\varphi^{(2)} + \cdots\end{array}
\end{equation}
Once the ans\"atze \eqref{kdv3} and \eqref{kdv4} are inserted back into
\eqref{kdv2}, the leading term is ${\cal O}(\epsilon^{1/2})$ and
simply yields $\varphi^{(1)}=0$ (and thus $v^{(1)}=0$). At next order
($\epsilon^{3/2}$) one obtains
\begin{equation}\label{kdv6}
\mathbb{K} \, \begin{pmatrix}
n_\xi^{(1)}\\[1mm] U_\xi^{(1)}\\[1mm]
\theta_\xi^{(1)}\\[1mm]\varphi^{(2)}\end{pmatrix}
=0\; ,
\end{equation}
where
\begin{equation}\label{kdv7}
\mathbb{K}=\begin{pmatrix}
c   & -\frac12 & k_0 & 0 \\
-g_1  & c      &  0   & 0 \\
k_0 & 0       & c   & -\Omega \\
0    & -k_0     &\Omega+g_2 & 0 \\
\end{pmatrix}
\; .
\end{equation}
Since $\det \mathbb{K} = 0$, Eq.~\eqref{kdv6} has non trivial
solutions. The kernel of $\mathbb{K}$ is one-dimensional; as a result,
the solution of Eq.~\eqref{kdv6} is of the form:
\begin{equation}\label{kdv8}
\begin{pmatrix}
 n^{(1)}_\xi \\[1mm] U^{(1)}_\xi \\[1mm] \theta^{(1)}_\xi \\[1mm] \varphi^{(2)}
\end{pmatrix} =
\begin{pmatrix}
	1 \\[1mm]\frac{g_1}{c} \\[1mm]
\frac{1}{c}\frac{k_0 g_1}{\Omega + g_2} \\[1mm]
\frac{k_0}{\Omega}
\frac{\Omega+g_1+g_2}{\Omega+g_2}
\end{pmatrix}
n^{(1)}_\xi
\equiv R\,n^{(1)}_\xi
\;.
\end{equation}
We also need (for later use) to determine the column vector $L$
such that
\begin{equation}\label{kdv9}
\mathbb{K}^t \, L =0 \Leftrightarrow L^t \, \mathbb{K} =0 \; ,
\end{equation}
This fixes
\begin{equation}\label{kdv10}
L^t \propto
\left(1,\frac{c}{g_1}, 0 ,
-\frac{k_0}{\Omega+g_2}\right)\; .
\end{equation}
At order $\epsilon^{5/2}$ we obtain
\begin{equation}\label{kdv11}
\mathbb{K} \, \begin{pmatrix}
n_\xi^{(2)}\\[1mm] U_\xi^{(2)}\\[1mm]\theta_\xi^{(2)}
\\[1mm]\varphi^{(3)}\end{pmatrix}
= \begin{pmatrix}
A_1\\[1mm] A_2\\[1mm]A_3\\[1mm]A_4\end{pmatrix}
\; ,
\end{equation}
where
\begin{equation}\label{kdv12}
\begin{split}
A_1=\;&
	n^{(1)}_\tau+ \tfrac12 [n^{(1)} U^{(1)}]_\xi -
k_0 [n^{(1)} \theta^{(1)}]_\xi \; ,\\
A_2=\;&	U^{(1)}_\tau + \tfrac12 U^{(1)} U^{(1)}_\xi - \Omega \,
\theta^{(1)} \theta^{(1)}_\xi \\
& + \tfrac12 n^{(1)}_{\xi\xi\xi}
+ k_0 v^{(2)}_\xi \; ,\\
A_3=\;& \theta^{(1)}_\tau+\tfrac12 U^{(1)}\theta_\xi^{(1)}
-\tfrac12 v_\xi^{(2)}\\
&+k_0\theta^{(1)}\theta_\xi^{(1)}+k_0 n^{(1)}n_\xi^{(1)}\; ,
 \\
A_4=\;& -g_2[n^{(1)} \theta^{(1)}]_\xi+
\tfrac12 \,\theta^{(1)}_{\xi\xi\xi}  -c \,v^{(2)}_\xi\; .
\end{split}
\end{equation}
Performing the substitution \eqref{kdv8}, we can express the system
\eqref{kdv11} in the following way:
\begin{equation}\label{kdv12-2}
\mathbb{K} \, \begin{pmatrix} n_\xi^{(2)}\\[1mm]
U_\xi^{(2)}\\[1mm]\theta_\xi^{(2)}\\[1mm]\varphi^{(3)}\end{pmatrix}
= C_\tau \, n_\tau^{(1)}
+C_3\,n^{(1)}_{\xi\xi\xi} + C_{\mathrm{nl}}\, n^{(1)}n^{(1)}_\xi \;,
\end{equation}
with
\begin{equation}\label{kdv12-3z}
C_\tau =\begin{pmatrix}
1 \\[1mm]\frac{g_1}{c} \\[1mm]
\frac{1}{c}\frac{k_0 g_1}{\Omega + g_2} \\[1mm]
0
\end{pmatrix} \; ,
\end{equation}
\begin{equation}\label{kdv12-3a}
C_{\mathrm{nl}}=
\begin{pmatrix}
 \frac{\sqrt{2} \sqrt{g_1 \left(-2 k_0^2+\Omega
+g_2\right)}}{\sqrt{\Omega +g_2}} \\
 g_1 \left(\frac{\Omega }{\Omega +g_2}+\frac{g_2}{-2 k_0^2+\Omega
+g_2}\right) \\
 k_0 \left(\frac{g_1 \left(2 k_0^2+\Omega +g_2\right)}{\left(\Omega
+g_2\right) \left(-2 k_0^2+\Omega +g_2\right)}+1\right) \\
 -\frac{2 \sqrt{2} \sqrt{g_1} g_2 k_0}{\sqrt{\Omega +g_2} \sqrt{-2
k_0^2+\Omega +g_2}} \\
\end{pmatrix}\; ,
\end{equation}
and
\begin{equation}\label{kdv12-3b}
C_3=
\begin{pmatrix}
 0 \\
 \frac{\left(\Omega +g_1+g_2\right) k_0^2}{\Omega \left(\Omega
+g_2\right)}-\frac{1}{2} \\
 -\frac{\left(\Omega +g_1+g_2\right) k_0}{2 \Omega \left(\Omega
+g_2\right)} \\
 \frac{k_0 \left(2 \left(\Omega +g_1+g_2\right) k_0^2-\left(\Omega
+g_2\right) \left(g_1+g_2\right)\right)}{\sqrt{2} \Omega \left(\Omega
+g_2\right){}^{3/2} \sqrt{\frac{-2 k_0^2+\Omega +g_2}{g_1}}} \\
\end{pmatrix}
\; .
\end{equation}

Left multiplication of Eq.~\eqref{kdv12-2} by $L^t$ gives
\begin{equation}\label{kdv13}
0 = (L^t\cdot C_\tau) \, n_\tau^{(1)} + (L^t\cdot C_{\mathrm{nl}}) \,
n^{(1)}n^{(1)}_\xi + (L^t\cdot C_3) \, n^{(1)}_{\xi\xi\xi} \;.
\end{equation}
Equation \eqref{kdv13} is a consistency condition: Eq.~\eqref{kdv11}
admits a solution only if the column vector $A$ is in the image space
of $\mathbb{K}$, which is implied by \eqref{kdv13} (we used the same
technique in Sec.~\ref{sec:upperbranch}, see Eqs.~\eqref{Lorder2},
\eqref{eq:Phi_rho} and \eqref{eq:GP1_a}).
Explicitly, Eq.~\eqref{kdv13} reads
\begin{equation}\label{kdv14}
n^{(1)}_\tau + \frac{3g_1}{4c}
\left(1 -
\frac{2 \,\Omega\, k_0^2}{\big(\Omega+g_2 \big)^2}\right)
n^{(1)}n^{(1)}_\xi-c_3\, n^{(1)}_{\xi\xi\xi} = 0 \;,
\end{equation}
where $c_3$ is the third order coefficient of the dispersion relation
\eqref{sol:disp}.  Going back to the original coordinates $x$ and $t$
and denoting $n'(x,t)=n(x,t)-1$, we obtain the KdV equation
\begin{equation}\label{kdv15}
n'_t + c\,n'_x + \gamma_1\,
n'\,n'_x-c_3\, n'_{xxx} = 0\; ,
\end{equation}
where
\begin{equation}\label{kdv16}
\gamma_1 = \frac{3g_1}{4c}
\left(1 -
\frac{2\,\Omega\, k_0^2}{\left(\Omega+g_2 \right)^2}\right)\; .
\end{equation}
Once the solution of Eq.~\eqref{kdv15} is
found, the other field variables can be obtained using relations
\eqref{kdv8} which we rewrite here for completeness in the final
notation:
\begin{equation}\label{kdv17}
\begin{split}
U(x,t)=\; & \frac{g_1}{c}\, n'(x,t)\; ,\\
\theta(x,t)=\; & -\frac{\pi}{2}+\frac{k_0g_1}{c\,(\Omega+g_2)}\, n'(x,t)\; ,\\
\varphi(x,t)=\; & \frac{k_0}{\Omega}
\frac{\Omega+g_1+g_2}{\Omega+g_2}\, n'_x(x,t)\;
.\end{split}
\end{equation}

Note that when $n'$ becomes of order of the nonlinear coefficient in
the KdV Eq.~\eqref{kdv15}, that is when $\gamma_1\sim |n'| \ll 1$,
the level of accuracy accepted here is not sufficient: the cubic
nonlinear terms ($\sim n^{\prime\,2} n'_x$)
neglected in the present treatment have
the same order of magnitude as the quadratic term in
Eq.~\eqref{kdv15}. In this limit we have to consider
the next order of approximation.

\subsection{Cubic nonlinearity: Gardner regime}\label{gardner_regime}

As advocated in Sec.~\ref{KdVregime}, cubic nonlinearities become
important when $\gamma_1\sim n'$ is small: their contributions can
therefore be calculated from the system \eqref{mod5} choosing
parameters such that $\gamma_1=0$. This is achieved when
\begin{equation}\label{gar1}
k_0=\frac{\Omega+g_2}{\sqrt{2 \Omega}} \;.
\end{equation}
For this choice of parameters, the sound velocity
\eqref{eq:sound_v} reads
\begin{equation}\label{cstar}
c=c^{\star} \equiv \sqrt{\frac{-g_1 g_2}{2\, \Omega}} \; .
\end{equation}
In this case the system can sustain long wavelength perturbations only
if $g_2<0$, that is if $\alpha_2>\alpha_1$, which we assume
henceforth (see however the discussion at the end of Sec. \ref{quartic}
and Appendix \ref{app2}).

In this regime the coordinates defined in \eqref{kdv1} are no longer
appropriate for the description of non-linear excitations. One should
instead perform the computations with the new coordinates:
\begin{equation}\label{gar2}
\xi=\epsilon(x-ct) \; , \quad\mbox{and}\quad \tau=\epsilon^3 t \; .
\end{equation}
Then the system \eqref{kdv2} rewrites:
\begin{subequations}\label{gar3}
\begin{align}
\epsilon^3 n_\tau=\;& \epsilon c\,n_\xi +
\epsilon k_0 \left[n \cos\theta\right]_\xi  \nonumber \\
-\;&
\tfrac12 \epsilon \big[n\left(U-v\cos\theta \right)\big]_\xi \;,
 \\
\epsilon^3 U_\tau  =\; &
\epsilon c\,U_\xi - \epsilon k_0 \, v_\xi \nonumber \\
-\;& \epsilon
 \left[\frac{\epsilon^2}{4} \frac{n_\xi^2}{n^2}
- \frac{\epsilon^2}{2} \frac{n_{\xi\xi}}{n} - \epsilon^2 \frac{\cot \theta}{2}
\frac{\left[n \, \theta_\xi \right]_\xi}{n}\right]_\xi \nonumber  \\
	-\;&\epsilon
\left[\frac{\epsilon^2\theta_\xi^2+U^2+v^2}{4}+ g_1 n \right]_\xi
 \nonumber  \\
+\;&\epsilon\theta_\xi\frac{\Omega \cos\theta}{\sin^2\theta} \cos\varphi
+\frac{\Omega}{\sin\theta} \, v\, \sin\varphi \;,
 \\
\epsilon^3\theta_\tau  =\; & \epsilon c \, \theta_\xi
- \Omega \, \sin\varphi \nonumber  \\
-\;& \tfrac12 \epsilon \left(U \theta_\xi
+ \frac{\left[n\,(v+2k_0)\sin\theta \right]_\xi}{n} \right) \;,
\\
\epsilon^3 v_\tau=\; & \epsilon c\,v_\xi - \epsilon k_0 \, U_\xi \nonumber \\
+\;& \epsilon
\left[\frac{\epsilon^2}{2 \sin \theta}
\frac{\left[n \, \theta_\xi \right]_\xi}{n}-\frac{U\,v}{2}
+ g_2 n
\cos\theta\right]_\xi \nonumber  \\
	+\;&\epsilon \theta_\xi\,\frac{\Omega \cos\varphi}{\sin^2\theta}
+ \Omega \, v\, \cot\theta\, \sin\varphi \;.
\end{align}
\end{subequations}
We perform a multiscale analysis using the ans\"atze \eqref{kdv3};
$U$ and $v$ now read:
\begin{equation}
U=\epsilon\, \Phi_\xi\; ,\quad\mbox{and}\quad v=\epsilon\, \varphi_\xi\; .
\end{equation}
The leading term in \eqref{gar3} is now ${\cal O}(\epsilon)$ and
reads, as previously, $\varphi^{(1)}=0$. The next order
${\cal O}(\epsilon^2)$ is described by the same system as in
Eq.~\eqref{kdv6}.  Substituting $U^{(1)}_\xi$, $\theta^{(1)}_\xi$ and
$\varphi^{(2)}$ by their expression in $n^{(1)}$ defined in
\eqref{kdv8}, the order ${\cal O}(\epsilon^3)$ then reads :
\begin{equation}\label{gar4}
\mathbb{K} \, \begin{pmatrix}
n_\xi^{(2)}\\[1mm] U_\xi^{(2)}
\\[1mm]\theta_\xi^{(2)}\\[1mm]\varphi^{(3)}\end{pmatrix}=
C_{\mathrm{nl}}  \, n^{(1)} n^{(1)}_\xi
\; ,
\end{equation}
where $C_{\mathrm{nl}}$ is defined in Eq.~\eqref{kdv12-3a}.  Since in
this subsection $k_0$ is fixed such that the non-linearity
($\gamma_1\propto L^t C_{\rm nl}$) in Eq.~\eqref{kdv15} cancels, the
choice of parameter \eqref{gar1} automatically implies $L^t \cdot
C_{\mathrm{nl}}=0$ and from Eq.~\eqref{gar4} one can only deduces that
\begin{equation}\label{gar5}
\begin{pmatrix}
n_\xi^{(2)}\\[1mm] U_\xi^{(2)}\\[1mm]\theta_\xi^{(2)}\\[1mm]\varphi^{(3)}
\end{pmatrix} =
\begin{pmatrix}
0 \\
0 \\
\frac{2\sqrt{g_1(-g_2)}}{\Omega+g_2} \\
\frac{g_1 \left(3 g_2 \Omega -g_2^2+2 \Omega ^2\right)-g_2
\left(g_2+\Omega \right){}^2}{\sqrt{2} g_2 \Omega \left(g_2+\Omega
\right)}
\end{pmatrix} n^{(1)} n^{(1)}_\xi \;.
\end{equation}
Eq.~\eqref{gar4} is thus not conclusive and the expansion at
order ${\cal O}(\epsilon^3)$ is not sufficient to describe the dynamic of
non-linear excitations.  At next order [${\cal O}(\epsilon^4)$], taking into
account the formula \eqref{gar5}, we obtain :
\begin{equation}\label{gar6}
\mathbb{K} \, \begin{pmatrix} n_\xi^{(3)}\\[1mm]
U_\xi^{(3)}\\[1mm]\theta_\xi^{(3)}\\[1mm]\varphi^{(4)}\end{pmatrix}
=
C_\tau n_\tau^{(1)}
+ D_{\mathrm{nl}}\, {n^{(1)}}^2n^{(1)}_\xi +C_3\,n^{(1)}_{\xi\xi\xi} \;,
\end{equation}
where $C_\tau$ and $C_3$ are defined in Eqs. \eqref{kdv12-3z},
\eqref{kdv12-3b} and
\begin{equation}\label{gar7}
D_{\mathrm{nl}} =
\begin{pmatrix}
 \frac{\Omega  \sqrt{-g_1^3 g_2}+\sqrt{-g_1 g_2} g_2
\left(g_1-6 g_2\right)}{2 \sqrt{2} \sqrt{\Omega } g_2^2} \\
 -\frac{3 \Omega  g_1}{\Omega +g_2} \\
 -\frac{2 g_2 \left(\Omega +g_2\right){}^2+g_1
\left(\Omega ^2-8 g_2 \Omega -5 g_2^2\right)}{2 \sqrt{2} \sqrt{\Omega } g_2
\left(\Omega +g_2\right)} \\
 \frac{\sqrt{-g_1 g_2} \left(-6 g_2^3-g_1 \left(2 \Omega -g_2\right)
\left(\Omega +g_2\right)\right)}{2 g_2^2 \left(\Omega +g_2\right)} \\
\end{pmatrix}\; .
\end{equation}
Left multiplication of Eq.~\eqref{gar6} by $L^t$ gives
\begin{equation}\label{gar8}
0 = (L^t\cdot C_\tau) \, n_\tau^{(1)} + (L^t\cdot D_{\mathrm{nl}}) \,
{n^{(1)}}^2n^{(1)}_\xi + (L^t\cdot C_3) \, n^{(1)}_{\xi\xi\xi} \;.
\end{equation}
which reads
\begin{equation}\label{gar9}
\begin{split}
0 = n_\tau^{(1)} +
\frac{3\,g_1}{8\,c^*\,|g_2|} \left(g_1 - \frac{4g_2^2}{\Omega+g_2} \right) \,
{n^{(1)}}^2n^{(1)}_\xi-c_3 \, n^{(1)}_{\xi\xi\xi} \;.
\end{split}
\end{equation}
Going back to the original coordinates $x$ and $t$ we obtain the
modified KdV (mKdV) equation
\begin{equation}\label{gar10}
n'_t+ c^*\,n'_x + \gamma_2\, n^{\prime\,2} n'_x -c_3 \, n'_{xxx}=0\; ,
\end{equation}
where
\begin{equation}\label{gar11}
\begin{split}
\gamma_2=\; & \frac{3\,g_1}{8\,c^*\,|g_2|}
\left(g_1 - \frac{4g_2^2}{\Omega+g_2} \right) \\
=\; &
\frac{3}{4\sqrt{2}} \sqrt{\frac{\Omega g_1}{-g_2^3}} \left(g_1 -
\frac{4g_2^2}{\Omega+g_2} \right)
\; .
\end{split}
\end{equation}
In the regime where $\gamma_1$ is not exactly zero, but of order
$n'$, we also have to take into account the quadratic nonlinearity of
Eq.~\eqref{kdv15}, which finally yields
\begin{equation}\label{gar12}
n'_t+ c\,n'_x + \gamma_1 n' n'_x
+ \gamma_2\, n^{\prime\,2} n'_x -c_3 \, n'_{xxx}=0\; .
\end{equation}
This is the Gardner equation describing the evolution of nonlinear
polarization pulses in a coherently coupled two-component condensate
in the limit where the parameters of the system are close to satisfy
the condition \eqref{gar1}. This can be considered as an intermediate
region where the quadratic and cubic nonlinearities make contributions
of the same order of magnitude in the wave dynamics.  In the limit of
very small $\gamma_1$, the quadratic nonlinearity effects can be
neglected, the nonlinear polarization waves are correctly described by
the modified KdV equation \eqref{gar10}. If instead $\gamma_1$ is
large, then the cubic nonlinearity effects are negligible and the
evolution of nonlinear polarization pulses is described by the KdV
equation \eqref{kdv15}.  Note that for consistency reasons, the value
of the sound velocity in \eqref{gar12} has to evaluated as not being
exactly equal to $c^*$, but has to include corrections $\propto
\gamma_1^2$.

Once the solution of the Gardner equation \eqref{gar12} has been
found, the other field variables can be expressed in terms of $n'$ by
the formulas \eqref{kdv17} with account of \eqref{gar1}.

\subsection{Quartic nonlinearity: generalized KdV
  equation}\label{quartic}

The parameters of the system can be chosen in such a way that not only
$\gamma_1$, but also $\gamma_2$ cancels. This is achieved for the
choice \eqref{gar1} with the additional constrain
\begin{equation}\label{quartic1}
g_1=\frac{4g_2^2}{\Omega+g_2}\; ,
\end{equation}
which ensures that $\gamma_2=0$. Note that for having a positive value
of $g_1$, one must have $\Omega+g_2>0$, but this condition is
automatically fulfilled in phase III [cf. the definition
\eqref{eq:sound_v} of the sound velocity]. Computations very
similar to the ones exposed in Secs. \ref{KdVregime} and
\ref{gardner_regime} now lead to a higher order mKdV equation
\begin{equation}\label{quartic2}
0 = n_\tau^{(1)} + \gamma_3\,
{n^{(1)}}^3 n^{(1)}_\xi -c_3 \, n^{(1)}_{\xi\xi\xi} \;,
\end{equation}
where
\begin{equation}\label{quartic3}
\gamma_3=\frac{5 \sqrt{2}\,
g_2^2 \,\Omega \left(\Omega-2 g_2 \right)\left(\Omega
+g_2\right)^{9/2} }{\sqrt{-g_2 \Omega }} \; .
\end{equation}

Finally we can write the general form which is able to account for
choices of parameters such that $\gamma_1\simeq 0$ and $\gamma_2\simeq
0$:
\begin{equation}\label{quartic4}
n'_t + C(n')n'_x-c_3\, n'_{xxx} = 0\; ,
\end{equation}
with
\begin{equation}\label{quartic5}
C(n')=c + \gamma_1\, n' + \gamma_2\, n^{\prime\,2} +
\gamma_3\, n^{\prime\,3} \; .
\end{equation}
Eq.~\eqref{quartic4} is known as a generalized KdV equation \cite{Tsu70}.

At this point, it might be helpful to remind the strategy followed in
the present section: we study excitations of the lower branch of the
spectrum, which, in the linear regime and in the long wave limit, are
described by Eq.~\eqref{eq17}.  In order to analyze how nonlinearity
affects these excitations, we consider the modifications of the
pulse propagation velocity induced by the nonlinear effects: $c \to C(n')$ with
an expansion of the form $C(n') = c + \sum_{\ell \ge 1} \,
\gamma_\ell\, {n'}^\ell$.  The multi-scale analysis consists in
rescaling the variable $(x,t)$ in the following way:
\begin{equation}\label{phil2}
\xi = \epsilon^a (x-c\,t) \quad\text{and}\quad
\tau = \epsilon^b t \;,
\end{equation}
transforming Eq.~\eqref{eq17} into
\begin{equation}\label{phil3}
\epsilon^b n'_\tau = \underset{\ell \ge 1}{\sum} \epsilon^{a+\ell} \gamma_\ell
\, {n'}^\ell\,n'_\xi + \epsilon^{3a} c_3\,n'_{\xi\xi\xi} \;.
\end{equation}
The analysis amounts to determine the coefficients $\gamma_\ell$; this
has been done in Eqs.~\eqref{kdv16}, \eqref{gar11} and
Eq.~\eqref{quartic3}. The first correction is $\ell=1$ :
\begin{equation}\label{phil4}
\epsilon^b n'_\tau = \epsilon^{a+1} \gamma_1\,
n'\,n'_\xi + \epsilon^{3a} c_3\,n'_{\xi\xi\xi} \;.
\end{equation}
The approach is coherent if all orders in $\epsilon$ in Eq.~\eqref{phil4}
are identical, i.e., if the stretched variables in \eqref{phil2}
are chosen with $b=a+1=3\,a \Rightarrow (a=1/2,b=3/2)$.

The parameter $\gamma_1(g_1,g_2,k_0,\Omega)$ can vanish or become small for a
particular value of $k_0$; the first order correction $\ell=1$ is then
no longer sufficient, and we must consider the correction $\ell=2$
which corresponds, by the same argument as the one used after
Eq.~\eqref{phil4}, to $(a=1,b=3)$; these are the exponents used in
Sec.~\ref{gardner_regime}.  If $\gamma_2(g_1,g_2,k_0,\Omega)$ is also
small, one has to consider the next order, as done in the beginning of
the present section. The different orders considered and their regime
of relevance are recalled in Table~\ref{table1}.

\begin{table}[h]
\begin{center}
\begin{tabular}{|l|l|}
  \hline
  non-linear equation & regime of relevance \\
  \hline
  $\ell=1$ : KdV &  phase III \\
  $\ell=2$ : Gardner & $\gamma_1\simeq 0$ and $g_2<0$ \\
  $\ell=3$ : generalized KdV &  $\gamma_2\simeq 0$ and $|g_2|<\Omega$ \\
  \hline
\end{tabular}
\end{center}
\caption{List of the different nonlinear equations describing the weakly
  nonlinear and weakly dispersive dynamics of excitations which, in the linear
  regime, pertain to the lower dispersion branch. The right column shows their
  successive regime of relevance of the equations. The conditions
  $\gamma_1=0$ and $\gamma_2=0$ are precisely defined in Eqs.~\eqref{gar1} and
  \eqref{quartic1}. Note that each row assumes that
  the regime of relevance of the upper rows is fulfilled.}
\label{table1}
\end{table}

It is appropriate to discuss if the different regimes identified in
table \ref{table1} can be reached with current days experimental
realization of spin-orbit coupled BECs. The references
\cite{spin-orbit-exp} consider the two states $|m_F = 0\rangle =
|\!\uparrow\,\rangle$ and $|m_F = -1\rangle = |\!\downarrow\,\rangle$
of a $^{87}$Rb Bose-Einstein condensate in the $F=1$ hyperfine
structure.  The $s$-wave scattering lengths are (in units of the Bohr
radius) $a_{\uparrow\uparrow}=101.41$ and $a_{\uparrow\downarrow} =
a_{\downarrow\downarrow} = 100.94$. Since
$a_{\uparrow\uparrow}-a_{\downarrow\downarrow} \ll \frac12 (
a_{\uparrow\uparrow} + a_{\downarrow\downarrow})$ the simplifying
assumption of a common value of the nonlinear coupling $\alpha_1
=\hbar\omega_\perp(a_{\uparrow\uparrow} + a_{\downarrow\downarrow})$
in \eqref{mod1a} is legitimate (in this expression $\omega_\perp$ is
the angular frequency corresponding to a tight harmonic radial
trapping which ensures a quasi-1D behavior of the condensate
\cite{reduc1D}).  Besides, if necessary, the present formalism can be
extended to take into account the fact that $a_{\uparrow\uparrow}$ and
$a_{\downarrow\downarrow}$ are not equal, see Ref.~\cite{Kam14}.  Note
that the recoil energy is typically $\frac12 k_0^2\sim 2$~kHz (it is
monitored by the wavelength and the relative angle of the Raman
lasers), whereas the interaction energy $\frac12 g_1=\frac12
(\alpha_1+\alpha_2)\rho_0$ is of order $(1\div5)$~kHz (it depends on
the value of the radial trapping frequency and on the linear atomic
density). 

One has $\alpha_2 =2 \hbar\omega_\perp a_{\uparrow\downarrow} <
\alpha_1$, hence $g_2>0$, and it seems from the discussion in the
beginning of section \ref{gardner_regime} that one can never reach the
interesting regime where $\gamma_1=0$ and where the nonlinear
modulations of an excitation formed in the lower branch is described
by Gardner equation. However, $g_2$ is small (since $\alpha_1$ and
$\alpha_2$ are so close) and, as explained in Appendix \ref{app2},
generalizing the present approach by taking into account the small
detuning $\delta$ from the Raman resonance---not considered in the
main text---one can show that, even with a positive $g_2$, it is
possible to reach a regime where the nonlinearity coefficient
$\gamma_1$ cancels by correctly fixing the value of $\delta$. However,
for keeping the discussion simple we will only consider here the case of a
small negative $g_2$. The more relevant case of a small positive $g_2$
in the presence of a small detuning is presented in Appendix
\ref{app2}. The main conclusions are similar in both cases.  Then, for
negative $g_2$, the condition \eqref{gar1} leads to $\Omega=2 k_0^2 -2
g_2 +{\cal O}(g_2^2)$, which corresponds to a value of the Raman
coupling frequency $\Omega$ typical in present days experiments. We
recall however that for the system to remain in the good side of the
boundary between phase III and phase II one needs to impose $\Omega>2
k_0^2 - g_2$, which is verified by the above choice of $\Omega$, but
not by a large extend. Hence, one can reach a regime where the lower
excitation branch is described by Gardner dynamics, but this is
obtained at the expense of getting close to the phase III--phase II
boundary. Away from this boundary, the lower branch has a KdV
dynamics.

\section{Conclusion}\label{conclusion}

In the present paper we have described how nonlinearity affects the
dynamics of elementary excitations of a coherently coupled
Bose-Einstein condensate. Excitations in the upper branch of the
spectrum display a modulational instability. As discussed in the text,
this instability is stabilized by a low wave--short wave resonance: the
momentum imparted by the wave train formed in the upper branch to
excitations in the lower branch has a stabilizing effect when it has
the same sign that the velocity of the wave train. We also showed that
the system can experience second harmonic generation, and that this
mechanism may be inhibited by symmetry effects (namely by the complete
separation between density and polarization modes which occurs when
$k_0=0$).

Excitations in the lower branch are stable. In the long wave length
limit they are affected by nonlinear effects in a manner which can
generically be described by KdV dynamics. For some specific
configuration of the system's parameters (close to the phase II--phase
III boundary) one has to use Gardner equation instead. It is
interesting to note that the Gardner regime is realized in the lower
branch which is a mode mainly corresponding to density waves: hence
the wide range of nonlinear excitations of Gardner's equation (see,
e.g., Ref.~\cite{Kam12}) can be generated by means of a simple scalar
potential, whereas for non-coherently coupled two component
condensates, where the Gardner regime is obtained for a polarization
mode \cite{Kam14}, this can be achieved only thanks to a polarization
potential \cite{Kam13}.

\begin{acknowledgments}
  We thank G. Martone and A. Recati for fruitful discussions. AMK
  thanks Laboratoire de Physique Th\'eorique et Mod\`eles Statistiques
  (Universit\'e Paris-Sud, Orsay) where this work was started, for
  kind hospitality.  This work was supported by the French ANR under
  grants n$^\circ$ ANR-11-IDEX-0003-02 (Inter-Labex grant QEAGE) and
  ANR-15-CE30-0017 (Haralab project).
\end{acknowledgments}

\appendix

\section{Solution of Eq.~\eqref{eq:Phi_rho_b}}\label{app1}

In this Appendix we briefly explain how the solution of
Eq.~\eqref{eq:Phi_rho_b} is obtained.
Let us assume that it is of the form
\begin{equation}\label{ansatz-thib}
\overline\Phi^{(1)}=W(k) \int^X\!\!\! \mathrm{d}X
  \left|\widetilde\theta^{(1)}\right|^2\; ,
\end{equation} where $W(k)$ is a constant (i.e., it depends on $k$, but not
on $(X,T_1,T_2)$).  One first remarks that
\begin{equation}\label{eq:etape}\begin{split}
\partial_{T_1} \left|\widetilde\theta^{(1)}\right|^2
& =  \widetilde\theta^{(1)*}\partial_{T_1}\widetilde\theta^{(1)} +  \cc\\
& \stackrel{\rm Eq. \eqref{eq-importante}}{=}
-\omega'_+(k)\, \widetilde\theta^{(1)*}\partial_{X}\widetilde\theta^{(1)} +  \cc\\
& = -\omega'_+(k)\, \partial_{X} \left|\widetilde\theta^{(1)}\right|^2\; .
\end{split}
\end{equation}
It follows from this result and from the ansatz \eqref{ansatz-thib} that
$\partial^2_{T_1}\overline\Phi^{(1)}=W(k)\,[\omega'_+(k)]^2
\partial_X |\widetilde\theta^{(1)}|^2$, and of course
$\partial^2_{X} \overline\Phi^{(1)} = W(k)\,
\partial_X |\widetilde\theta^{(1)}|^2$ [this is a direct consequence
of \eqref{ansatz-thib}]. Hence
\begin{equation}
\label{eq:Phi_rho_bbis}
\begin{split}
& \partial_{T_1}^2 \overline\Phi^{(1)} - c^2 \, \partial_{X}^2
\overline\Phi^{(1)} =\\
&  W(k)\,\bigg([\omega'_+(k)]^2-c^2\bigg)\,
\partial_X \left|\widetilde\theta^{(1)}\right|^2 \; .
\end{split}
\end{equation}
Equating the r.h.s. of this expression to the r.h.s. of
\eqref{eq:Phi_rho_b} determines the value of $W(k)$ as given in
Eq.~\eqref{sol:Phi_rho_b}.

\section{Taking into account a small detuning from
the Raman  resonance}\label{app2}

In this Appendix we rapidly present the treatment of the lower
excitation branch for a spin-orbit coupled condensate ($k_0\ne 0$)
in the case where the system experiences a finite
detuning $\hbar\, \delta$ from the Raman resonance. The single particle
Hamiltonian $H_0$ \eqref{mod1b} has now an additional contribution:
$\frac12 \hbar \,\delta\, \sigma_z$. The system \eqref{mod5} is not
modified, except for Eq.~\eqref{mod5d} which now reads
\begin{equation}\label{b1}
\begin{split}
\varphi_t=&\frac{1}{2\sin\theta}\frac{(\rho\,\theta_x)_x}{\rho}
-\tfrac12 \Phi_x(\varphi_x+2k_0)+ \\
& (\alpha_1-\alpha_2)\rho \cos\theta-\Omega\,\cos\varphi\cot\theta +
\delta\; .
\end{split}
\end{equation}
The ground state value of the fields is no longer given by
Eq.~\eqref{mod7}. One has now
\begin{equation}\label{b2}
\Xi^{(0)}(t)=\begin{pmatrix} 1 \\ 2 k_1 x -2\mu t \\ \theta_0 \\ 0\end{pmatrix}
\; ,
\end{equation}
where $k_1$ is a variational parameter. Minimizing the energy per
particle one obtains \cite{Li14}
\begin{equation}\label{eqk1}
k_1=k_0 \cos\theta_0 \; .
\end{equation}
The same result can be obtained in a different manner, by using dynamical
arguments: one keeps $k_1$ as a free parameter, and one studies the linear
excitations of the system. By demanding that the system is dynamically
stable, i.e., that the frequency of elementary excitations
remains real, one obtains the result \eqref{eqk1}.

\begin{figure}[t]
\includegraphics[width=0.99\linewidth]{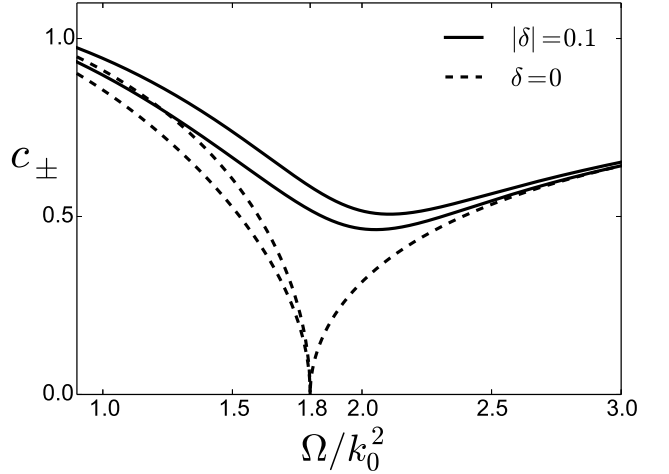}
\caption{Sound velocities $c_{\pm}$ as a function of $\Omega$. The
  system's parameter are are $k_0=1$, $g_1=2.2$ and $g_2=0.2$. When
  $\delta=0$, the transition from phase III to phase II occurs at
  $\Omega=2 k_0^2-g_2=1.8$, and above $\Omega=1.8$ one is in phase III
  with a single sound velocity. When $\delta\neq 0$ one can show that
  $c_+(-\delta)=c_-(\delta)$ and $c_-(-\delta)=c_+(\delta)$. This is
  the reason why the sign of $\delta$ is not specified in the
  figure.}\label{fig.sound-omega}
\end{figure}

In the
presence of a finite $\delta$, the ground state value $\theta_0$ is no
longer exactly equal to $-\pi/2$ as in \eqref{mod7} and $\mu$ is not given by
Eq.~\eqref{mod6}. Instead one has [from \eqref{b1} and \eqref{mod5b}]
\begin{equation}\label{b3}
\left(2k_0^2-g_2\right) \cos \theta_0 + \Omega \cot \theta_0 = \delta \; ,
\end{equation}
and
\begin{equation}\label{b4}
\mu=\frac{k_0^2}{2}\left(1+\cos^2\theta_0\right)+
\frac{g_1}{2} +\frac{\Omega}{2\, \sin \theta_0} \; .
\end{equation}
Eq.~\eqref{b3} determines the value of $\theta_0$. Depending on
the system's parameters it has either four or two
solutions. In the first case, only one corresponds to the minimum of
the energy per particle (the other is the maximum) and the system can
be considered to be in the single minimum phase III. In the second
case there are two non equivalent minima and the system is in phase
II. In the regime where $k_0^2$ is larger than $g_2/2$ and where
$\Omega>0$, one can show that the boundary between these two regimes
corresponds to
\begin{equation}\label{b5}
\left(2\,k_0^2-g_2\right)^{2/3} = \Omega^{2/3}+ |\delta|^{2/3} \; .
\end{equation}
In the case $\delta=0$ the solution of \eqref{b3} is $\theta_0=-\pi/2$
if $\Omega>2 k_0^2-g_2$ and \eqref{b5} corresponds to the standard
transition line between phases II and III which is reproduced in
Figure~\ref{fig.phases}. It is important to stress that in the
presence of a finite detuning $\delta$ the second order phase
transition from phase III to phase II strictly speaking disappears
because the system does not cross any phase transition line when
$\Omega$ varies \cite{remII-III}. For instance, the velocity of sound
vanishes at the transition region when $\delta=0$
(cf. Eq.~\eqref{eq:sound_v}), whereas it remains finite when
$\delta\neq 0$ (cf. Fig. \ref{fig.sound-omega}). Also, when
$\delta\neq 0$, even in what has been identified above as the single
minimum phase, the system has a small spin polarization and condensates
into a state with a small but finite momentum.

The matrix $\mathbb{K}$ of Eq. \eqref{kdv7} now reads
\begin{equation}\label{b6}
\mathbb{K}=\begin{pmatrix}
c   & -\frac12 & -k_0 \sin_0 & 0 \\
-g_1  & c -k_0 \cos_0
  &  \Omega \cos_0\sin^{-2}_0 & 0 \\
-k_0 \sin_0 & 0       & c-2k_0 \cos_0   & -\Omega \\
g_2 \cos_0    & -k_0 &\Omega \sin^{-2}_0 -g_2  \sin_0 & 0 \\
\end{pmatrix}
\; ,
\end{equation}
where, for gaining space, we have written $\sin_0$ and $\cos_0$
instead of $\sin\theta_0$ and $\cos\theta_0$. In formula \eqref{b6}
the sound velocity $c$ is not given by \eqref{eq:sound_v}: it now
depends on $\delta$. It can be determined through the computation of
the dispersion relation in the system, or more simply just by imposing
the cancellation of $\det \mathbb{K}$. For non zero $\delta$ the ground
state breaks Galilean invariance and in our 1D configuration one
obtains two velocities of sound, one for each direction of
propagation. The first one -- denoted as $c_+$ -- corresponds to
waves propagating in the same direction as the ground state (for which
$U_x^{(0)}=2 k_1$) and the other ($c_-$) propagates in the opposite
direction. A typical case is displayed in
Fig. \ref{fig.sound-omega}. Note that this figure is interrupted at low
values of $\Omega$ in order to prevent the system to get into phase I.

\begin{figure}[h]
\includegraphics[width=0.99\linewidth]{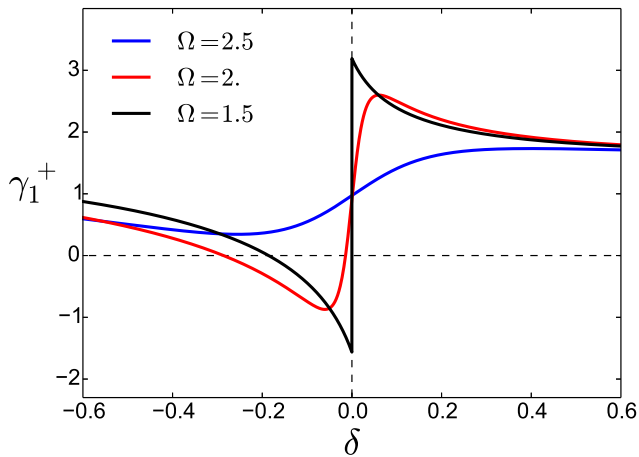}
\caption{(Color online). $\gamma_1^+$ as a function of $\delta$ for
  several values of $\Omega$. The system's parameter are $k_0=1$,
  $g_1=2.2$ and $g_2=0.2$. The curve is discontinuous when $\Omega<2
  k_0^2-g_2=1.8$ because in this case the system meets a first order
  phase transition at $\delta=0$.  For the present choice of
  parameters $\gamma_1^+$ can be canceled by changing the value of
  $\delta$ when $\Omega=1.5$ or 2.5, but not when
  $\Omega=2.5$.}\label{fig.gamma1-delta}
\end{figure}

Following the procedure exposed in section \ref{KdVregime} one can
determine the form of the KdV equation which describes how long
wavelength excitations propagating along the lower branch of the spectrum
are affected by nonlinearity when $\delta\neq 0$. In this case the
nonlinear parameter $\gamma_1$ is different from the value given by
expression \eqref{kdv16} (which corresponds to the $\delta=0$
case). We do not write the explicit form of $\gamma_1$ when
$\delta\neq 0$ because it is too cumbersome. Instead, we rather plot
$\gamma_1$ as a function of $\Omega$ for different values of $\delta$
in Figure~\ref{fig.gamma1-delta}.
As for the sound velocity, the value of the nonlinear coefficient
depends of the direction of propagation of the wave. We denote as
$\gamma_1^+$ ($\gamma_1^-$) the value of $\gamma_1$ corresponding to
wave-trains propagating in the same (the opposite) direction than the
momentum of the ground state. One has the symmetry relation
$\gamma_1^+(-\delta)=\gamma_1^-(\delta)$. When $\Omega<2 k_0^2-g_2$
the nonlinear coefficient is discontinuous at $\delta=0$, as $\theta_0$
and $k_1$ are, and this corresponds to the crossing of the first order
transition line in the plane $(\Omega,\delta)$ \cite{remII-III}.

One sees in the figure that there exist values of $\delta$ for which
the nonlinear coefficient cancels even for $\Omega > 2 k_0^2-g_2$,
provided $\Omega$ is not too large. It is important to notice
that this cancellation of $\gamma_1$ is obtained for a positive $g_2$,
contrarily to what occurs when $\delta=0$. Note however
that for $\Omega=2.5$ the
non-linear coefficient cannot be canceled by imposing a finite value
of the detuning $\delta$. When $\gamma_1$ cancels, the effective
nonlinear dynamics of the system is no longer described by a KdV
equation, but rather by a Gardner equation.


\begin{thebibliography}{99}

\bibitem{Wya97} C. J. Myatt, E. A. Burt, R. W. Ghrist, E. A. Cornell,
and C. E. Wieman, Phys. Rev. Lett. {\bf 78}, 586 (1997).

\bibitem{domains} J. Stenger, S. Inouye, D.M. Stamper-Kurn,
  H.-J. Miesner, A.P.  Chikkatur, and W. Ketterle, Nature (London)
  {\bf 396}, 345 (1998); H.-J. Miesner, D. M. Stamper-Kurn,
  J. Stenger, S. Inouye, A. P. Chikkatur, and W. Ketterle
  Phys. Rev. Lett. {\bf 82}, 2228 (1999); M. R. Matthews,
  B. P. Anderson, P. C. Haljan, D. S. Hall, C. E. Wieman, and
  E. A. Cornell, Phys. Rev. Lett. {\bf 83}, 2498 (1999); Z. Dutton, M.
  Budde, C. Slowe, L. Vestergaard Hau, Science {\bf 293}, 663 (2001);
  N. S. Ginsberg, J. Brand, and L. V. Hau, Phys. Rev. Lett. {\bf 94},
  040403 (2005); L. E. Sadler, J. M. Higbie, S. R. Leslie,
  M. Vengalattore, and D. M.  Stamper-Kurn, Nature (London) {\bf 443},
  312 (2006); K. C. Wright, L. S. Leslie, A. Hansen, and
  N. P. Bigelow, Phys. Rev. Lett. {\bf 102}, 030405 (2009);
  D. M. Weld, P. Medley, H.  Miyake, D. Hucul, D. E. Pritchard, and
  W. Ketterle, Phys. Rev. Lett. {\bf 103}, 245301 (2009);
  L. S. Leslie, A. Hansen, K. C. Wright, B. M. Deutsch, and
  N. P. Bigelow, Phys. Rev. Lett. {\bf 103}, 250401 (2009).

\bibitem{Josephson} M.-S. Chang, Q. Qin, W. Zhang, L. You and
  M. S. Chapman, Nat. Phys. {\bf 1}, 111 (2005); T. Zibold,
  E. Nicklas, C. Gross, and M. K. Oberthaler, Phys. Rev. Lett. {\bf 105},
  204101 (2010).

\bibitem{intrication} P. B\"ohi, M. F. Riedel, J. Hoffrogge, J. Reichel,
T. W. H\"ansch, and P. Treutlein, Nat. Phys. {\bf 5}, 592 (2009);
C. Gross, T. Zibold, E. Nicklas, J. Est\`eve, and
M. K. Oberthaler, Nature (London) {\bf 464}, 1165 (2010);
M. F. Riedel, P. B\"ohi, Y. Li, T. W. H\"ansch,
  A.  Sinatra, and P. Treutlein, Nature (London) {\bf 464}, 1170 (2010);
  C. Gross, H. Strobel, E. Nicklas, T. Zibold, N. Bar-Gill,
  G. Kurizki, and M. K. Oberthaler, Nature (London) {\bf 480}, 219 (2011).

\bibitem{impurity} S. Palzer, C. Zipkes, C. Sias, and M.K\"ohl,
  Phys. Rev. Lett. {\bf 103}, 150601 (2009); T. Fukuhara, A. Kantian,
  M. Endres, M. Cheneau, P. Schau\ss, S. Hild, D. Bellem,
  U. Schollw\"ock, T. Giamarchi, C. Gross, I.  Bloch, and S. Kuhr,
  Nat. Phys. {\bf 9}, 235 (2013).

\bibitem{Bea13} S. Beattie, S. Moulder, R. J. Fletcher, and Z. Hadzibabic
Phys. Rev. Lett. {\bf 110}, 025301 (2013).

\bibitem{gauge} Y.-J. Lin, R. L. Compton, A. R. Perry, W. D. Phillips,
  J. V. Porto, and I. B. Spielman, Phys. Rev. Lett. {\bf 102}, 130401
  (2009); Z. Fu, P. Wang, S. Chai, L. Huang, and J. Zhang,
Phys. Rev. A {\bf 84}, 043609 (2011); M. W. Ray, E. Ruokokoski, S. Kandel,
M. M\"ott\"onen, and D. S. Hall, Nature {\bf 505}, 657 (2014).

\bibitem{spin-orbit-exp} Y.-J. Lin, K. Jim\'enez-Garc\'ia, and
  I. B. Spielman, Nature (London) {\bf 471}, 83 (2011);
J.-Y. Zhang, S.-C. Ji, Z. Chen, L. Zhang, Z.-D. Du, B. Yan, G.-S. Pan,
B. Zhao, Y.-J. Deng, H. Zhai, S. Chen, and J.-W. Pan,
Phys. Rev. Lett. {\bf 109}, 115301 (2012); S.-C. Ji, J.-Y.
Zhang, L. Zhang, Z.-D. Du, W. Zheng, Y.-J. Deng, H. Zhai, S. Chen,
and J.-W. Pan, Nat. Phys. {\bf 10}, 314 (2014).

\bibitem{Leb12} L. J. LeBlanc, K. Jim\'enez-Garc\'ia, R. A. Williams,
  M. C. Beeler, A. R. Perry, W. D. Phillips, and I. B. Spielman,
  Proc. Natl. Acad. Sci. USA {\bf 109}, 10811 (2012).

\bibitem{Qu13} C. Qu, C. Hamner, M. Gong, C. Zhang, and P. Engels,
Phys. Rev. A {\bf 88}, 021604(R) (2013); L. J. LeBlanc, M. C. Beeler, K.
Jim\'enez-Garc\'ia, A. R. Perry, S. Sugawa, R. A. Williams, and I. B.
Spielman, New J. Phys. {\bf 15}, 073011 (2013).

\bibitem{Bee13} M. C. Beeler, R. A. Williams, K. Jim\'enez-Garc\'ia,
L. J. LeBlanc, A. R. Perry, and I. B. Spielman, Nature (London) {\bf 498},
201 (2013).

\bibitem{Ols14} A. J. Olson, S.-J. Wang, R. J. Niffenegger, C.-H. Li,
  C. H. Greene, and Y. P. Chen, Phys. Rev. A {\bf 90}, 013616 (2014).

\bibitem{Ham14} C. Hamner, C. Qu, Y. Zhang, J. Chang, M. Gong, C. Zhang, and
P. Engels, Nat. Comm. {\bf 5}, 4023 (2014).

\bibitem{roton} M. A. Khamehchi, Y. Zhang, C. Hamner, T. Busch, and P.
Engels, Phys. Rev. A {\bf 90}, 063624 (2014); S.-C. Ji,
L. Zhang, X.-T. Xu, Z. Wu, Y. Deng, S. Chen, and J.-W. Pan,
Phys. Rev. Lett. {\bf 114}, 105301 (2015).

\bibitem{two1} M. R. Matthews, D. S. Hall, D. S. Jin, J. R. Ensher,
  C. E. Wieman, E. A. Cornell, F. Dalfovo, C. Minniti, and
  S. Stringari, Phys. Rev. Lett. {\bf 81}, 243 (1998); D. S. Hall,
  M. R. Matthews, C. E. Wieman, and E. A. Cornell,
  Phys. Rev. Lett. {\bf 81}, 1543 (1998).

\bibitem{two2} K. C. Wright, L. S. Leslie, and N. P. Bigelow,
  Phys. Rev. A {\bf 77}, 041601(R) (2008); K. C. Wright, L. S. Leslie,
  and N. P. Bigelow, Phys. Rev. A {\bf 78}, 053412 (2008); J. Higbie
  and D. M. Stamper-Kurn, Phys. Rev. Lett. {\bf 88}, 090401 (2002).

\bibitem{GS1} P. B. Blakie, R. J. Ballagh, and C. W. Gardiner,
  J. Opt. B: Quantum Semiclass. Opt. {\bf 1}, 378 (1999); C. P. Search
  and P. R. Berman, Phys. Rev. A {\bf 63}, 043612 (2001);
  P. Tommasini, E. J. V. de Passos, A. F. R. de Toledo Piza,
  M. S. Hussein, and E. Timmermans, Phys. Rev. A {\bf 67}, 023606
  (2003); C. Lee, W. Hai, L. Shi, and K. Gao, Phys. Rev. A {\bf 69},
  033611 (2004); M. Abad and A. Recati, Eur. Phys. J. D {\bf 67}, 148
  (2013).

\bibitem{GS2} C. Wang, C. Gao, C.-M. Jian, and H. Zhai,
  Phys. Rev. Lett. {\bf 105}, 160403 (2010); T.-L. Ho and S. Zhang,
  Phys. Rev. Lett. {\bf 107}, 150403 (2011); Y. Zhang, L. Mao, and
  C. Zhang, Phys. Rev. Lett. {\bf 108}, 035302 (2012); W. Zheng and
  Z. Li, Phys. Rev. A {\bf 85}, 053607 (2012); Y. Li,
  L. P. Pitaevskii, and S. Stringari, Phys. Rev. Lett. {\bf 108},
  225301 (2012); G. I. Martone, Y. Li, L. P. Pitaevskii, and
  S. Stringari, Phys. Rev. A {\bf 86}, 063621 (2012); Y. Zhang,
  G. Chen, and C. Zhang, Scientific Reports {\bf 3}, 1937 (2013);
  W. Zheng, Z.-Q. Yu, X. Cui, and H. Zhai, J. Phys. B:
  At. Mol. Opt. Phys. {\bf 46}, 134007 (2013); Y. Li, G. I. Martone,
  L. P. Pitaevskii, and S. Stringari, Phys. Rev. Lett. {\bf 110},
  235302 (2013).

\bibitem{NL1} D. T. Son and M. A. Stephanov, Phys. Rev. A {\bf 65},
  063621 (2002); J. J. Garc\'ia-Ripoll, V. M. P\'erez-Garc\'ia, and
  F. Sols, Phys. Rev. A {\bf 66}, 021602(R) (2002); K. Kasamatsu,
  M. Tsubota, and M. Ueda, Phys. Rev. Lett. {\bf 93}, 250406 (2004);
  B. Deconinck, P. G. Kevrekidis, H. E. Nistazakis, and
  D. J. Frantzeskakis, Phys. Rev. A {\bf 70}, 063605 (2004);
  V. A. Brazhnyi and V. V. Konotop, Phys. Rev. E {\bf 72}, 026616
  (2005); I. M. Merhasin, B. A. Malomed, and R. Driben; J. Phys. B:
  At. Mol. Opt. Phys. {\bf 38}, 877 (2005); K. Nakamura, A. Kohi, H.
  Yamasaki, V. M. P\'erez-Garc\'ia, and V. V. Konotop, EPL {\bf 80},
  50005 (2007).

\bibitem{NL2} M. Merkl, A. Jacob, F. E. Zimmer, P. \"Ohberg, and
  L. Santos, Phys. Rev. Lett. {\bf 104}, 073603 (2010); X.-Q. Xu and
  J. H. Han, Phys. Rev. Lett. {\bf 107}, 200401 (2011); T. Kawakami,
  T. Mizushima, M. Nitta, and K. Machida, Phys. Rev. Lett. {\bf 109},
  015301 (2012); O. Fialko, J. Brand, and U. Z\"ulicke, Phys. Rev. A
  {\bf 85}, 051605(R) (2012); V. Achilleos, D. J. Frantzeskakis,
  P. G. Kevrekidis, and D.E. Pelinovsky, Phys. Rev. Lett. {\bf 110},
  264101 (2013); V. Achilleos, J. Stockhofe, P. G. Kevrekidis,
  D. J. Frantzeskakis, and P. Schmelcher, EPL {\bf 103}, 20002 (2013);

\bibitem{Kam15} C. Polymilis, K. Hizanidis, and D. J. Frantzeskakis,
  Phys. Rev.  E {\bf 58}, 1112 (1998); P.L. Christiansen,
  J. C. Eilbeck,V. Z. Enolskii, and N. A. Kostov, Proc. R. Soc. London
  A {\bf 456}, 2263 (2000); C. Eilbeck, V. Z. Enolskii, and
  N. A. Kostov, J. Math. Phys.  {\bf 41}, 8236 (2000);
  A. M. Kamchatnov and V. V. Sokolov, Phys. Rev. A {\bf 91}, 043621
  (2015).

\bibitem{Li14} Y. Li, G. I. Martone, and S. Stringari, chapter in
  volume III of the {\it Annual Review of Cold Atoms and Molecules},
  ed. K. W. Madison, K. Bongs, L. D. Carr, A. M. Ray and H. Zhai,
  p. 201 (World Scientific, Singapore, 2015).

\bibitem{ktu-2005} K. Kasamatsu, M. Tsubota, and M. Ueda,
  Phys. Rev. A {\bf 71}, 043611 (2005).

\bibitem{rem_theta} The value $\theta_0=+\pi/2$ also gives equal
  densities of the two components, but is unstable when $\Omega>0$.

\bibitem{Jef82} A. Jeffrey and T. Kawahara, {\it Asymptotic Methods in
    Nonlinear Wave Theory}, Pitman, London, 1982.

\bibitem{Tan83} T. Taniuti and K. Nishihara, {\it
Nonlinear Waves}, Pitman, Boston, 1983.

\bibitem{New85} A. C. Newell, {\it Solitons in Mathematics and Physics},
Society for Industrial and Applied Mathematics, Philadelphia, 1985.

\bibitem{Kam00} A. M. Kamchatnov, {\it Nonlinear Periodic Waves and
Their Modulations---An Introductory Course} (World Scientific,
Singapore, 2000).

\bibitem{Abl11} M. J. Ablowitz, {\it Nonlinear Dispersive Waves,
    Asymptotic Analysis and Solitons}, Cambridge University Press,
  Cambridge, 2011.

\bibitem{Rembar} The bar is not a complex conjugate. The complex
  conjugate is denoted in this work with a superscript $^*$.

\bibitem{remshg} This is not always true: in some instances the
  determinant of $\mathbb{M}_2$ cancels. This occurs in the presence
  of second harmonic generation, as discussed in section
  \ref{final-upper} [after Eqs. \eqref{shg1} and \eqref{shg2}].

\bibitem{rem1} We use the simple property that the image space of a
  given matrix $\mathbb{M}$ is orthogonal---in the sense of the usual
  scalar product---to the kernel of $\mathbb{M}^t$, to which $L$
  belongs by \eqref{consistant}. This property is easily demonstrated in
  $\mathbb{R}^n$: Let $C$ be a column vector in the image space of a
  $n\times n$ matrix $\mathbb{M}$; this means that there exists a
  column vector $V$ such that $C=\mathbb{M} V$. Let $L$ be a column vector
  in the kernel of $\mathbb{M}^t$. This means that $\mathbb{M}^t L=0
  \Leftrightarrow L^t\, \mathbb{M} =0$. Then it is clear that the
  scalar product $L^t\, C=0$ (since it reads $L^t\, \mathbb{M} C$).

\bibitem{rembackxt} From Eqs. \eqref{derivatives} and
  \eqref{eq-importante} one gets
  $\partial_{T_2}\widetilde{\theta}^{(1)} = \eps^{-2}\partial_t
  \widetilde{\theta}^{(1)} -\eps^{-1}\partial_{T_1}
  \widetilde{\theta}^{(1)} = \eps^{-2}\partial_t
  \widetilde{\theta}^{(1)} +\eps^{-1}\omega_+'(k)\partial_X
  \widetilde{\theta}^{(1)}$.

\bibitem{Zak09} V. E. Zakharov and L. A. Ostrovsky, Physica D {\bf
    238}, 540 (2009).

\bibitem{mi-ol} V. V. Konotop and M. Salerno Phys. Rev. A {\bf 65},
  021602(R) (2002); A. Smerzi, A. Trombettoni, P. G. Kevrekidis, and
  A. R. Bishop, Phys. Rev. Lett. {\bf 89}, 170402 (2002); B. Wu,
  Q. Niu, New J. Phys. {\bf 5}, 104 (2003); M. Machholm, C.J. Pethick,
  H. Smith, Phys. Rev. A {\bf 67}, 053613 (2003); M. Cristiani,
  O. Morsch, N. Malossi, M. Jona-Lasinio, M. Anderlini, E. Courtade,
  and E. Arimondo, Opt. Express {\bf 12}, 4 (2004); L. Fallani, L. De
  Sarlo, J. E. Lye, M. Modugno, R. Saers, C. Fort, and M. Inguscio
  Phys. Rev. Lett. {\bf 93}, 140406 (2004); L. De Sarlo, L. Fallani,
  J. E. Lye, M. Modugno, R. Saers, C. Fort, and M. Inguscio
  Phys. Rev. A {\bf 72}, 013603 (2005).

\bibitem{mi-ab} C. K. Law, C. M. Chan, P. T. Leung, and M.-C. Chu,
  Phys. Rev. A {\bf 63}, 063612 (2001); M. A. Hoefer, J. J. Chang,
  C. Hamner, and P. Engels Phys. Rev. A {\bf 84}, 041605(R) (2011);
  C. Hamner, J. J. Chang, P. Engels, and M. A. Hoefer
  Phys. Rev. Lett. {\bf 106}, 065302 (2011).

\bibitem{Benj-Feir} T. Brooke Benjamin and J. E. Feir, J. Fluid Mech.
  {\bf 27}, 417 (1967).

\bibitem{LWSW} D. J. Benney, Stud. Appl. Math. {\bf 56}, 81 (1976);
R. H. J. Grimshaw, Stud. Appl. Math. {\bf 56}, 241 (1977).

\bibitem{Kiv98} Y. S. Kivshar and B. Luther-Davies, Phys. Rep. {\bf 298}, 81
(1998).

\bibitem{Pet02} C. J. Pethick and H. Smith, {\it Bose-Einstein
    condensation in Dilute Gases}, Cambridge University Press,
  Cambridge, 2002.

\bibitem{Pit03} L. Pitaevskii and S. Stringari, {\it Bose-Einstein
    Condensation}, Clarendon Press, Oxford, 2003.

\bibitem{Tsu70} M. Tsutsumi, T. Mukasa, and R. Iino, Proc. Japan Acad.
{\bf 46}, 921 (1970).

\bibitem{reduc1D} M. Olshanii, Phys. Rev. Lett. {\bf 81}, 938 (1998);
  A. D. Jackson, G. M. Kavoulakis, and C. J. Pethick, Phys. Rev.  A
  {\bf 58}, 2417 (1998); P. Leboeuf and N. Pavloff, Phys. Rev. A {\bf
    64}, 033602 (2001).

\bibitem{Kam14} A. M. Kamchatnov, Y. V. Kartashov, P.-\'E. Larr\'e and
  N. Pavloff, Phys. Rev. A {\bf 89}, 033618 (2014).

\bibitem{Kam12} A. M. Kamchatnov, Y.-H. Kuo, T.-C. Lin, T.-L. Horng, S.-C.
Gou, R. Clift, G. A. El, and R. H. J. Grimshaw, Phys. Rev. E {\bf 86},
036605 (2012).

\bibitem{Kam13} A. M. Kamchatnov and Y. V. Kartashov,
  Phys. Rev. Lett. {\bf 111}, 140402 (2013).

\bibitem{remII-III} The situation is similar to the
  magnetization transition in a pure ferromagnet: in the absence of
  external field the system has a line of first order transition which
  terminates with a critical point. For fixed finite external magnetic
  field, one does not cross this line by varying temperature. It
  can be crossed at fixed temperature by varying the external field.
  In our case, the equivalent line corresponds in the
  $(\Omega,\delta)$ plane to the segment $\delta=0$, $|\Omega|\le
  2k_0^2-g_2$ which is a line of first order phase transition ending
  with a critical point at $\delta=0$ and $|\Omega|= 2k_0^2-g_2$. This
  line cannot be crossed at fixed non zero $\delta$ by changing
  $\Omega$, but can be at fixed $\Omega$ by changing $\delta$.

\end{thebibliography}
\end{document}